\documentclass[pdftex,dvipsnames,usenames,aps,superscriptaddress,twocolumn,amsmath,showkeys]{revtex4}

\usepackage{float}
\usepackage{dcolumn}
\usepackage{graphicx}
\usepackage{color}
\usepackage{amssymb}
\usepackage{bold-extra}
\usepackage{hyperref}
\usepackage{verbatim}
\usepackage[utf8]{inputenc}
\usepackage{lipsum}

\newcolumntype{d}[1]{D{.}{.}{#1} }
\newcolumntype{h}[1]{D{-}{-}{#1} }

\def\vB{{\bf{v}\times\bf{B}}}
\def\vvB{{\bf{v}\times\left(\bf{v}\times\bf{B}\right)}}

\def\beq{\begin{equation}}
\def\eeq{\end{equation} }
\def\bea{\begin{eqnarray}}
\def\eea{\end{eqnarray}}

\def\figref#1{Fig.~\ref{fig:#1}}
\def\figlab#1{\label{fig:#1}}  
\def\eqref#1{Eq.~(\ref{eq:#1})}

\def\eqlab#1{\label{eq:#1}}
\def\tabref#1{Table~\ref{tab:#1}}
\def\tablab#1{\label{tab:#1}}  
\newcommand*{\secref}[1]{Section~\ref{sec:#1}}
\newcommand*{\seclab}[1]{\label{sec:#1}}
\newcommand{\Omit}[1]{}
\newcommand{\Olaf}{\color[named]{Purple}}
\newcommand{\RED}{\color[named]{black}}
\newcommand{\BLUE}{\color[named]{Blue}}

\newcommand{\pragati}[1]{{\marginpar{\BLUE pragati}\RED\bf #1}}
\newcommand{\os}[1]{{\marginpar{\Olaf Olaf}\Olaf\bf #1}}

\def\Xmax{X_{\rm max}}
\def\Xrh{X_{\rm z}}
\def\Prog#1{} 

\def\KVI{Kapteyn Institute, University of Groningen, Groningen, The Netherlands}
\def\AIVUB{Astrophysical Institute, Vrije Universiteit Brussel, Pleinlaan 2, 1050 Brussels, Belgium}
\def\VUB{Vrije Universiteit Brussel, Dienst ELEM, Brussels, Belgium}
\def\NIKHEF{Nikhef, Science Park Amsterdam, Amsterdam, The Netherlands}
\def\IMAPP{Department of Astrophysics/IMAPP, Radboud University Nijmegen, Nijmegen, The Netherlands}

\def\ASTRON{Netherlands Institute for Radio Astronomy (ASTRON), Dwingeloo, The Netherlands}

\def\KIT{Institut f\"ur Astroteilchenphysik, KIT, P.O. Box 3640, 76021, Karlsruhe, Germany}
\def\FUW{Particles and Fundamental Interactions Division,Institute of Experimental Physics, University of Warsaw}
\def\CTU{Physics Education Department, School of Education, Can Tho University, Campus II, 3/2 Street, Ninh Kieu District, Can Tho City 94000, Vietnam}
\def\ECAP{Erlangen Center for Astroparticle Physics (ECAP), Friedrich-Alexander-Universit\"at Erlangen-N\"urnberg,
Nikolaus-Fiebiger-Straße 2, 91058 Erlangen, Germany}
\def \DESY {Deutsches Elektronen-Synchrotron DESY, Platanenallee 6, 15738 Zeuthen, Germany}

\def\KHALIFA {Khalifa University,  P.O. Box 127788, Abu Dhabi, United Arab Emirates}

\begin{document}

\title{Reconstructing Air Shower Parameters with MGMR3D}

\author{P.~Mitra} \email[]{pmitra@fuw.edu.pl} \affiliation{\FUW} \affiliation{\AIVUB} 
\author{O.~Scholten}  \email[]{o.scholten@rug.nl} \affiliation{\KVI}   \affiliation{\VUB}
\author{T.~N.~G.~Trinh} \affiliation{\CTU}
\author{S.~Buitink} \affiliation{\AIVUB} \affiliation{\IMAPP}
\author{J.~Bhavani} \affiliation{\KHALIFA}
\author {A.~Corstanje} \affiliation{\AIVUB} \affiliation{\IMAPP}
\author{M.~Desmet} \affiliation{\AIVUB}
\author{H.~Falcke} \affiliation{\IMAPP} \affiliation{\NIKHEF}
\author {B.~M.~Hare} \affiliation{\KVI} \affiliation{\VUB}
\author{J.~R.~H\"orandel} \affiliation{\IMAPP} \affiliation{\VUB}
\author{T.~Huege} \affiliation{\KIT} \affiliation{\AIVUB}
\author{N.~Karastathis} \affiliation{\KIT}
\author{G.~K.~Krampah} \affiliation{\AIVUB}
\author {K.~Mulrey} \affiliation{\IMAPP} \affiliation{\VUB}
\author{A.~Nelles} \affiliation{\DESY} \affiliation{\ECAP}
\author{H.~Pandya} \affiliation{\AIVUB}
\author{S.~Thoudam} \affiliation{\KHALIFA}

\author{K.~D.~de Vries} \affiliation{\VUB}
\author{S.~ter~Veen} \affiliation{\ASTRON}


\date{\today}

\begin{abstract}

Measuring the radio emission from cosmic ray particle cascades has proven to be a very efficient method to determine their properties such as the mass composition. Efficient modeling of the radio emission from air showers is crucial in order to extract the cosmic ray  physics parameters from the measured radio emission.  
MGMR3D is a fast semi-analytic code that calculates the complete radio footprint, i.e.\ intensity, polarization, and pulse shapes, for a parametrized shower-current density and can be used in a chi-square optimization to fit a given radio data. It is many orders of magnitude faster than its Monte Carlo counterparts.  We provide a detailed comparative study of MGMR3D to Monte Carlo simulations, where, with  improved  parametrizations, the shower maximum $\Xmax$ is found to have very strong agreement with a small dependency on the incoming zenith angle of the shower. Another interesting feature we observe with MGMR3D is sensitivity to the shape of the longitudinal profile in addition to $\Xmax$. This is achieved by probing the  distinguishable radio footprint produced by a shower having a different longitudinal profile than usual.
Furthermore, for the first time, we show the results of reconstructing shower parameters for LOFAR data using MGMR3D, and
obtaining a $\Xmax$ resolution of 22 g/cm$^2$ and energy resolution of 19\%.
\end{abstract}


\maketitle


\section{Introduction}

 When a high-energy cosmic particle impinges on the atmosphere of Earth, it creates an extensive air shower (EAS). The electrons and positrons in the plasma cloud at the shower front drift in opposite directions due to the Lorentz force caused by the geomagnetic field. Due to this acceleration by an Earth’s magnetic ﬁeld and deceleration in interactions with air molecules a time varying transverse current is created. 
 This varying current emits radio waves~\cite{olaf_geosync}  where the intensity pattern on the ground, the intensity footprint, depends on the variation of the current with height. There is another subdominant contribution to the radiation from the excess of negative charge ~ accumulated at the shower front, known as the 'Askaryan effect' \cite{askaryan1,askaryan2}. The penetration depth where the particle number reaches its maximum, $\Xmax$, strongly depends on the specifics of the first interaction, which strongly correlates with the mass of cosmic ray primary. Different values of $\Xmax$ result in differences in the longitudinal variation of the currents which is reflected in the intensity of the radio footprint. Thus $\Xmax$ can be reconstructed on the basis of the footprint which allows for a determination of the mass composition of cosmic rays~\cite{nature}.

The modeling of radio emission from EAS is generally performed with either microscopic or macroscopic formalisms.
In a microscopic formalism the emission is calculated for each particle as obtained from a Monte Carlo simulation of the EAS. The coherence of the signals emerges naturally in this approach.
ZHAires\cite{zharies} and CoREAS\cite{Huege:2013vt} are the two most commonly used microscopic codes.

MGMR \cite{olaf_geosync},  EVA \cite{krijn_cherenkov,krijn_eva} and their latest successor MGMR3D \cite{mgmr3d} are examples of macroscopic codes.
In this framework, the radiation field is derived from the Li\'enard-Wiechert potential~\cite{jacksonbook} where the  four-current is  parametrized. The amplitude of the four-current is explicitly split into the  charge component driving the charge excess emission and the transverse drift current generating the geomagnetic emission.
One advantage of MGMR3D is that it is computationally inexpensive and
produces radio profiles  about four orders of magnitude faster than  the  Monte Carlo simulations. Another advantage is that it is fully deterministic in the sense that one can have control over the outputs by choosing exact shower parameters like  the shower maximum and shape parameters of the longitudinal profile, contrary to the inherent randomness in Monte Carlo simulations. For these reasons, MGMR3D can be used to fit a reference radio footprint and  obtain the corresponding  longitudinal shower parameters  that best reproduce the given profile through minimization techniques. There are other, more phenomenological approaches emerging like template synthesis\cite{david_thesis}, radio morphing\cite{anna_radiomorph}, that also allow a fast calculation of the radio footprint.

 In MGMR3D the charge-current cloud of the air shower is parametrized which necessarily approximates its full complexity. 
In particular, the dependence on the energy of the particles forming this cloud is ignored, however, as the important particles in this cloud are relativistic, this is thought to be a reasonable approximation. In a prior publication ~\cite{mgmr3d}, the parametrization and the foundation of the MGMR3D framework were introduced. 
In this follow-up work, we further investigate the performance of MGMR3D on ensembles of air showers and have refined the parametrization in an extensive comparative study with CoREAS. Most significantly, we have used MGMR3D to re-analyze measured data obtained with LOFAR. The MGMR3D-based analysis reproduces, within statistical significance, the results of an earlier analysis based on microscopic CoREAS calculations. MGMR3D offers thus a very CPU-efficient alternative to existing approaches for extracting shower parameters like $\Xmax$ from the radio footprint, and thus composition, of the original cosmic rays. 
Notably, MGMR3D is also a strong tool to map atmospheric electric fields under thunderstorms. In a separate publication \cite{Trinh:2022} a detailed study is presented of using MGMR3D for reconstructing atmospheric electric fields during thunderstorm conditions from the radio footprint of air showers.


This article is structured as follows- In \secref{Modeling} we describe the improved modeling of the radiation profile. In \ref{sec:stokes_obs} and \ref{sec:Compare}, comparisons between CoREAS  with MGMR3D shower profiles are demonstrated, and the details of the results of fitting $\Xmax$. We also present a correction formula to obtain the correct zenith angle dependency for $\Xmax$ as compared to CoREAS calculations. Such a correction is necessary since the penetration depth for which the coherent transverse current is maximal generally differs from the penetration depth for which the number of charged particles is maximal, $\Xmax$.
We also report a study suggesting a strong correlation between showers with nonstandard shapes of longitudinal profiles and the fit quality of MGMR3D. This indicates a novel future prospect of extracting shower parameters regarding the shape of the longitudinal profile,  in addition to $\Xmax$, with radio technique using MGMR3D. This will in the end help gather a better understanding of the mass composition and hadronic models. In Section \secref{lofardata} we have shown the results of reconstructing $\Xmax$ using MGMR3D on measured LOFAR cosmic ray data and compare to the existing $\Xmax$ reconstructed with LOFAR analysis method, as well as the reconstruction of  shower core and energy.


\section{Modeling Radio emission from EAS}\seclab{Modeling}

The charge and current distributions that drive the radio emission from an EAS is expressed as a four-current $j^\mu(t,x,y,h)$ where $\mu=0$ denotes the time (charge) components, and $\mu=x,y,z$ denote the space (current) components.
The retarded Li\'{e}nard-Wiechert potential for an observer at ($t_o,x_o,y_o,z_o$) in the shower plane with the retarded time $t_r$ is
\beq
A^\mu(t_o,\vec{x_o})=\int d^3 \vec{x}'\,\frac{j^\mu(t_r,\vec{x}')}{\cal D} \;,
\label{L-W}
\eeq
where the retarded distance is
\beq
{\cal D}= n\sqrt{(-\beta t_o +h)^2 + (1-\beta^2 n^2)d^2}
 \label{denom} \;,
\eeq
where the distance between the observer and the point of impact of the core of the air shower is denoted by $d$, and the index of refraction is denoted by $n$.

Since for a cosmic-ray air shower the particles are concentrated in a relatively flat pancake-like structure moving with relativistic speeds, the four current is parametrized as
\beq
j^\mu(t,x,y,h)=\frac{w(r)}{r} \, f(h,r) \, J^\mu(t) \;.
\eqlab{DefCloud}
\eeq
The  term  $w(r) / r$ in \eqref{DefCloud} is the radial description of the plasma cloud, the second term $f(h,r)$ is the current density of the shower front. These two are normalised such that $J^\mu(t)$ is the charge and current for a fixed time integrated  over the complete plasma of the EAS. The radial dependence of the transverse
current is parametrized as
\begin{equation}
 w(r)= N_w \zeta \left(\zeta + 1\right)^{-2.5},
\end{equation}
with $\zeta=r/R_0$. 
The function $w(r) \times r$ corresponds  to  the NKG function \cite{nkg} for a fixed shower age $s=2$ \cite{mgmr3d}. These parametrizations were studied and optimized by comparing to the results of CONEX-MC\cite{krijn_eva}. The definition of  $R_0$ is similar to the Moli\`ere radius,  but not the same  as in this context it is a scaling parameter that describes the radial current profile and thus is referred to as radiation radius. 
In the original formulation of MGMR3D the radiation radius was taken to be a constant. We observed that the optimum value for $R_0$ depends on the distance from $\Xmax$ to the shower core ($D_{\Xmax}$), while fitting $R_0$ for different showers.  We find that for distances smaller than 5~km $R_0$ is proportional to distance and reaches saturation with $R_0=50$~m for larger distances, independent of zenith angle. This is shown in \figref{mol_radius}. This  linear dependency at smaller $D_{\Xmax}$ is now included in MGMR3D as $R_0= 10 \, D_{\Xmax}$.

\begin{figure}[t]
\includegraphics[width=1.1\columnwidth]{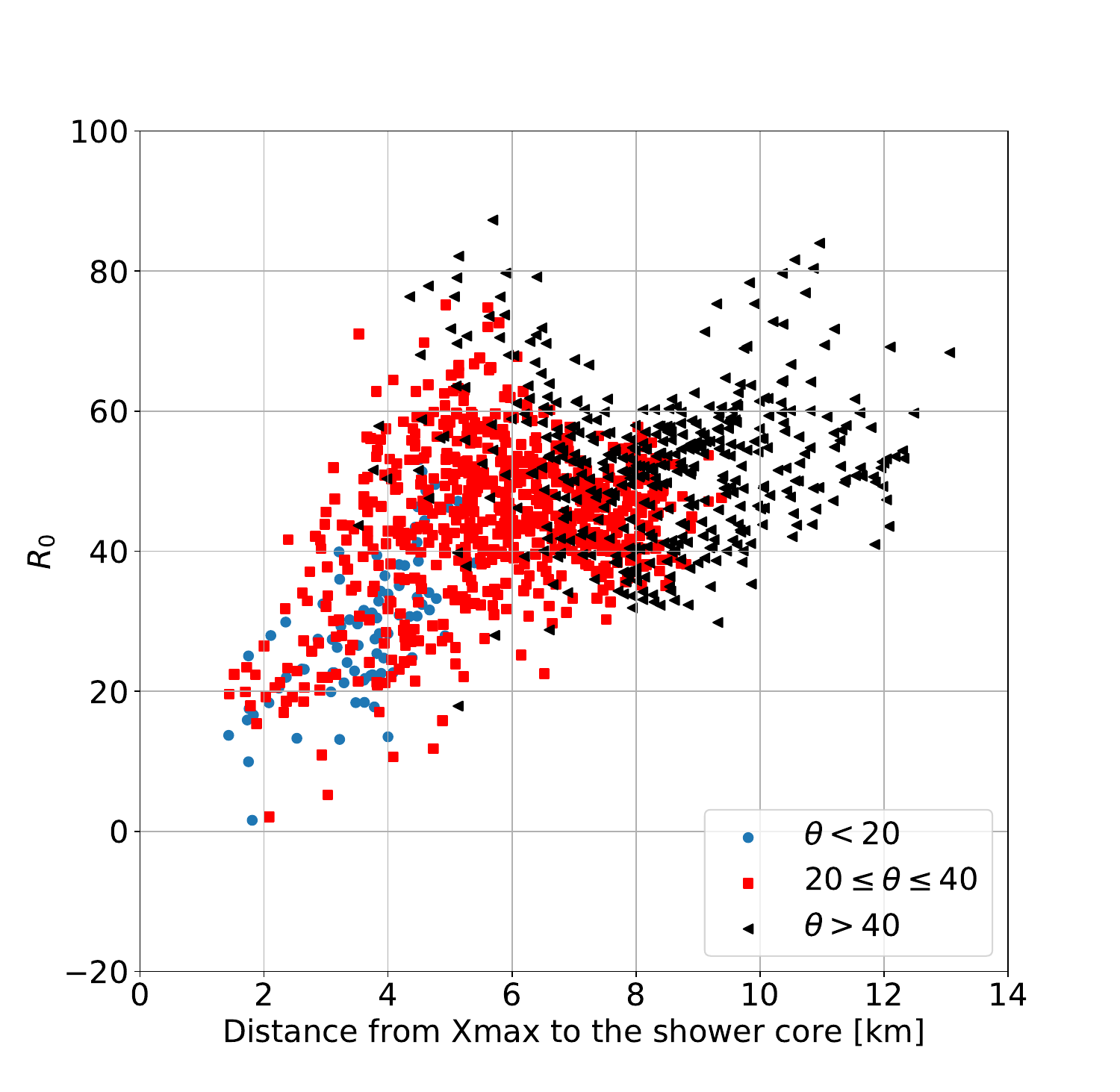}
\caption{Radiation radius as a function of distance of $\Xmax$ from shower core, obtained from comparing CoREAS showers to MGMR3D. Different colors show different zenith bands.  
}
\figlab{mol_radius}
\end{figure}


The current density at a distance $h$ behind the shower front is parametrized as
\begin{equation}
 f(h,r)=N_f \frac{\eta}{e^{\sqrt{\eta}+1}}.
\end{equation}
where $N_f$ is a normalisation constant. The parameter $\lambda$, folded in as $\eta=h/ \lambda$, accounts for the pancake thickness scaling and has a radial dependence. The radial dependence of the pancake thickness is described in a way that it is constant near the shower axis and increases linearly at distances away from the shower axis where particles tend to have less energy and thus lag behind. The parametrizations for the radial and pancake function were also studied and optimized with comparison to the results of CONEX-MC \cite{krijn_eva}.

The functions $w$ and $f$ that depends on the distance to the shower axis are normalized according to $\int_0^\infty  w(r)\, dr=1$ and $\int_0^\infty f(h,r)\, dh=1\; \forall\, r$.

\subsection{Parametrization of the currents}\seclab{ParamCurr}

The original parametrization of the charge cloud in MGMR3D, as described in \cite{mgmr3d}, was based on CONEX-MC simulations~\cite{krijn_eva}. There were however important inconsistencies in the extracted shower parameters as well as in the observed radiation profile, when compared to CoREAS results. A reason behind these differences  is  that in the parametrization the energy distribution of the particles in the shower is not taken into account. Parameters like the drift velocity and charge excess are strongly dependent on the energy range of particles used to predict these averaged quantities.  To mitigate these issues we revisit the parametrizations in this work by comparing the results of MGMR3D and CoREAS calculations for an ensemble of air showers. This leads to improved parametrizations, in particular for modeling the drift velocity (cf. \secref{ParamCurr}) and for the longitudinal profile of the current, $J^\mu$ in  \eqref{DefCloud}. The details of the comparison between MGMR3D and CoREAS are presented in \secref{Compare}.

 \subsubsection{Transverse current}

The transverse current is given by,

\beq
\vec{J}_\perp(t_s)=N_c(X_z) \, \vec{u}_\perp(\Xrh),
\eeq
where the transverse drift velocity is denoted as $\vec{u}_\perp(\Xrh)$,$N_c$ is the number of charged particles at depth ($X_z$).
It should be noted that the penetration depth for $\Xmax$ is only indirectly related to the penetration depth of maximum transverse current, since the factor between the two, the drift velocity, depends on air density as well as the mean energy of the particles in the shower.

The drift velocity  increases with increasing forces
acting on the charges. This becomes particularly important for large electric fields in thunderstorm clouds, and special treatment is required so that the
particles do not exceed the speed of light~\cite{gia1}. The transverse drift $\vec{u}_\perp(\Xrh)$ is therefore  expressed as \beq
\vec{u}_\perp(\Xrh)=\frac{c\vec{v}}{\sqrt{1+v^2/v_0^2}},
\eeq
where the parameter $v_0$  is adjusted to the value 0.2, and $v$ is taken proportional to the Lorentz force.

In the original parametrization used in  \cite{mgmr3d} no dependence on air density was assumed in the parametrization of the drift velocity. We noted that a $\sqrt{\rho}$ scaling was necessary to obtain agreement with the results of the CoREAS calculation. 
We thus updated the formula for the drift velocity to read
\beq
\vec{v}(X)= \frac{c\, \vec{F}_\perp}{F_t} \times \frac{a_t+1}{\frac{\Xmax-X_t}{\Xrh-X_t} + a_t} \times \sqrt{\frac{\rho(\Xmax)}{\rho(\Xrh)}}
\eqlab{Def-v}
\eeq
with  $ X_t=50$~g/cm$^2$, $F_\beta=250$~keV/c, and $a_t=3$.
$\vec{F}_\perp$ is the total transverse force acting on the particles, and for the air showers when no thunderstorm is present it only consists of  the Lorentz force, $\vec{F_{\perp}}=e\vec{v}_s \times\vec{B}$, where $\vec{v_s}$ is the velocity of the shower front, $e$ is the elementary charge, and $\vec{B}$ is  Earth's magnetic filed. The second factor  in \eqref{Def-v} takes into account the fact that the drift velocity depends on the penetration depth in the atmosphere, accounting for the changing mean energy of the shower particles. 
 It is good to mention that this parametrization becomes less accurate for the highest zenith angles, where an additional dependence on emission height is seen. This correction is not yet included in the code, which should therefore be used with caution when studying highly inclined showers above 60 degrees zenith angle. For the study reported in this article both simulated and recorded showers are well below this limit.

The physical interpretation  of the $\sqrt{\rho}$ scaling is not trivial. Interestingly,  the drift velocity has the same form  as the terminal velocity due to the macroscopic drag force acting opposite to the relative motion of any object moving in a fluid. The drag force of air  is proportional to the square of the speed of the object. For a falling object in air the terminal velocity can be reached when the force due to gravity balances the drag force
\begin{equation}
    mg=F_{D}= \frac{1}{2}\rho C A v^2 ,
\end{equation}
with $C,A,v$ being the drag coefficient, area of the object and terminal velocity respectively. Solving for $v$ results
\begin{equation}
    v=\sqrt{(2mg/\rho CA)}.
\end{equation}
The result can  be generalized to situations where the object is accelerated by other forces. In the case of the electron drift velocity that would be the Lorentz force. 
The equivalent of the drag force is actually due to the many elastic collisions of the relativistic electron in the shower front with neutral air molecules. A relativistic electron in the shower lives roughly a microsecond (300 meters)
before being stopped in a hard inelastic collision. Within
that microsecond, the electron actually undergoes more
than a million elastic collisions with particles in the air. 
While this provides an intuitive understanding of the $\rho^{-1/2}$ scaling, the assumption that an electron plasma experiences the same drag as a macroscopic object is of course not easily justified. It is worth mentioning that in \cite{anna_radiomorph} a similar density dependence  on the electric field amplitude of radio pulse  was  reported in a study for radio morphing method.

\begin{figure}[t]
\includegraphics[width=0.45\textwidth]{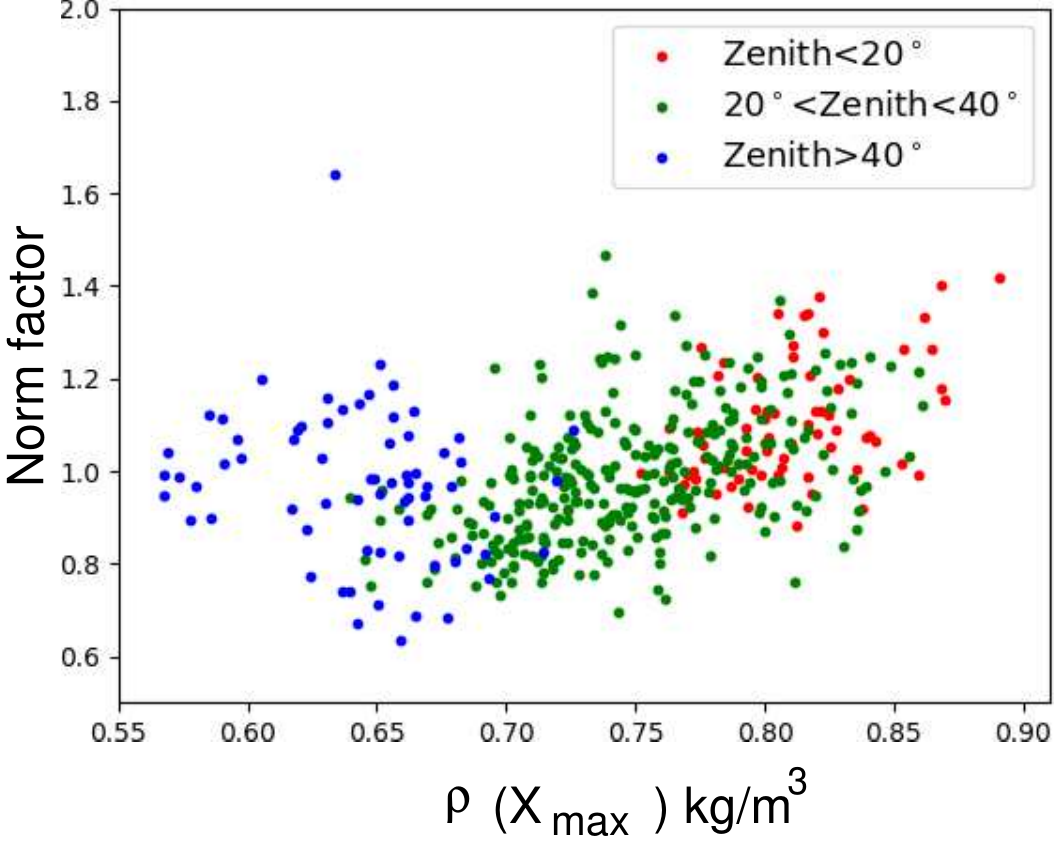}
\caption{Normalisation factor as a function of the density at $\Xmax$, obtained from MGMR3D by comparing to CoREAS showers. Different colors show different zenith angle bands.}
\figlab{normvsdensity}
\end{figure}

\subsubsection{Charge excess}

The charge excess  in the shower is given as, $J_Q(z)= e \, N_{c}(\Xrh)  \, \rho_c(\Xrh)$ where $e$ is the charge of the electron and the  proportionality factor is $\rho_c(\Xrh)$ defined in the most recent form,
\begin{align}
\rho_c(\Xrh) & = J^0_Q\, \frac{3 \Xrh- \Xmax -X_c}{\Xmax +\Xrh -X_c} & \notag \\
\times & \left(1 - e^{-\frac{\Xrh-X_c}{2(\Xmax-X_c)}}\right) \frac{\rho(\Xmax)}{\rho_c} \sqrt{\frac{\rho(\Xrh)}{\rho_c}}
\eqlab{Def-chxcurr}
\end{align}
where $J^0_Q$ is a normalisation constant, $\rho_c=0.06$~g/cm$^3$, and $ X_c=50$~g/cm$^2$. The first two factors in \eqref{Def-chxcurr} are inspired by comparing to the results of CONEX-MC simulations  including simulations for highly inclined showers with zenith $\>$ 65 degrees.
The last term including the square root dependency on density is inspired from the  treatment of transverse current in \eqref{Def-v}. \\


\subsection{Parametrization of the longitudinal profile}\seclab{ParamLP}

There are two common ways to parametrize  longitudinal profile, the number of charged particles at a depth $\Xrh$, one is the  Gaisser-Hillas formula~\cite{Gaisser:1977}, the other is the $R,L$ formula in \cite{rl_theory},

\beq
N^{\rm{R-L}}_{c}(\Xrh)= N_{\rm{max}} \times \left(1 - \frac{R}{L} (\Xmax -\Xrh)\right)^{R^{-2}} e^\frac{\Xmax - \Xrh}{ R \, L}
\eqlab{Def-Nc-RL}
\eeq
\beq
N^{\rm{G-H}}_{c}(\Xrh)= N_{\rm{max}} \times \left(\frac{ \Xrh-X_0}{\Xmax -X_0}\right)^\frac{\Xmax -X_0}{ \Lambda} e^\frac{\Xmax - \Xrh}{\Lambda}
\eqlab{Def-Nc-GH}
\eeq
where the number of particles at the shower maximum, $N_{\rm{max}}$ is taken proportional to the energy of the cosmic ray,
\beq
N_{\rm{max}}= N_E \, E_{cr}\;. \eqlab{norm}
\eeq
The constant $N_E$ is used as a norm factor when fitting the results of MGMR3D to data.
The main difference between the two  parametrizations  is that the parameters in \eqref{Def-Nc-GH} are  related to the  physics of the shower such as the depth of first interaction while $R$ and $L$ in \eqref{Def-Nc-RL} relate more directly to the rise and fall of the distribution \cite{shape_RL}. These more general parametrizations provide the option to study effects of the longitudinal shape parameters other than $\Xmax$ on the radio footprint \cite{Mitra:PhD}.
In principle, either of these parametrizations  can be used to describe the longitudinal profiles in MGMR3D. We have used \eqref{Def-Nc-RL} throughout this analysis.

The intensity of the radio pulse depends on the energy of the cosmic ray which is treated as a normalisation factor, a proxy for the air shower energy, in MGMR3D 
 when a $\chi^2$ fit to data is performed.
This normalisation factor was introduced in \eqref{norm}. 
Thus, when fitting the radio footprint as generated by CoREAS simulations for showers with a fixed energy, the normalisation factor should be constant, barring shower-to-shower fluctuations. In \figref{normvsdensity}, we indeed show this is approximately constant, for showers at various zenith angles. These values also have a global normalisation which is constant for all showers. 


\section{Stokes parameters as observables} \seclab{stokes_obs}
We investigate the radio footprint of an air shower using Stokes parameters since these capture the complete polarization structure of the radio pulse. Because the objective of the present work is to develop a scheme for data interpretation, we construct the Stokes parameters specific for the LOFAR frequency band, between 30 -- 80~MHz band.

The Stokes parameters can be expressed in terms of the
complex  observable ${\cal E}_{i}=E_{i} + i\hat{E}_{i}$, where ${E}_{i}$ is the 
electric field component  in  $\hat{e}_\vB$ and $\hat{e}_\vvB$ directions which are by construction perpendicular to the propagation direction of the shower, and $\hat{E}_{i}$ is its Hilbert transformation~\cite{pim_radio} (in arbitrary units), as
\begin{eqnarray}
I&=&\frac{1}{ N} \sum_0^{n-1} \left( |{\cal E}|^2_{i,\vB} + |{\cal E}|^2_{i,\vvB} \right) \nonumber\\
Q&=&\frac{1}{ N} \sum_0^{n-1} \left( |{\cal E}|^2_{i,\vB} - |{\cal E}|^2_{i,\vvB} \right) \nonumber\\
U +iV&=&\frac{2}{ N} \sum_0^{n-1} \left( {\cal E}_{i,\vB} \;  {\cal E}_{i,\vvB}^* \right) \;.
\label{Stokes}
\end{eqnarray}

We sum over the entire signal trace while calculating the values from CoREAS simulations.
The linear-polarization angle with the $\vB$-axis, $\psi$, can be calculated directly from the Stokes parameters as $\psi=\frac{1}{2} \tan^{-1} (U/Q)$. The relative amount of circular polarization is given by $V/I$ and it  can be interpreted due to a time lag between the peak of
the charge excess and transverse current pulses \cite{olaf_circpol}.

\subsection{Noise-error estimate on Stokes parameters}

MGMR3D performs a fit of the input radio profile through a Levenberg-Marquardt
minimization procedure~\cite{nl2sol}, that is based on a steepest descent method. The reduced $\chi^2$ of
the fit is defined as 
\beq
\chi^2=\frac{1}{N_{\rm{ndf}}} \, \sum_{{a,f}}\frac{(f_c^a-f_m^a)^2}{\sigma_{f^a}^2}
\eqlab{errdef}
\eeq
where $f_c^a$, and $f_m^a$ are the different Stokes parameters calculated with CoREAS  and MGMR3D respectively for antenna $a$, $N_{ndf}$
is the number of degrees of freedom, and $\sigma_f^a$ is the error on the  Stokes parameter.


\textcolor{black} {It is important to note that when we are performing a model-to-model comparison here, the numerator in \eqref{errdef}
does not have a noise contribution and the $\chi^2$ can be $<<$ 1. For the sake of clarity we refer this as $\tilde{\chi}^2$  throughout this paper to distinguish from standard $\chi^2$.}

In the present calculations, we calculated $\sigma_f$ for the comparison with CoREAS as 
\bea
\sigma_I^2 &=&   \frac{\Delta t} {2} \left( c  \epsilon \frac{2}{N} \sigma_n\, I + \frac{2}{N}\sigma_n^2\right) =   \frac{\Delta_t} {2} \left(\frac{2 \sigma_n}{N}(c \epsilon \, I_0+\sigma_n)\right) \nonumber \\
\sigma_Q^2 &=&  \frac{\Delta t} {2} \left( c \epsilon \,\frac{2}{N} \sigma_n\, I + \frac{2}{N}\sigma_n^2 \right) \nonumber \\
\sigma_U^2 &=& \sigma_V^2=  \frac{\Delta t} {2}\left(c \epsilon \,\frac{2}{N} \sigma_n\, I + \frac{2}{N}\sigma_n^2\right) \;,
\eqlab{errormodel}
\eea
where $N$ is the length of the trace and $\sigma_n$ is the noise fluence per sample,  $c$, $\epsilon$ are the natural constants - velocity of light and permittivity of air in S.I. units, and $\Delta t$ is the width of the time bins.

For measured cosmic ray data the value of the noise level $\sigma_{n}$  is obtained from measuring a time window of the recorded signal trace where  no significant signal is present. In the case of MGMR3D, the value is chosen such that it is a close representation of the measurement. the value is shown in \tabref{mytable1}. 

\section{Comparison to CoREAS simulations}\seclab{Compare}

\begin{table}
\caption{List of fixed parameters in MGMR3D.  OBSDIST$\_$DIM is the grid dimension used for the calculation of antenna distance to the shower
axis, J$_{0Q}$ is the charge excess normalisation factor, and $\sigma_n$ is the noise level. R and L are the shape parameters of the longitudinal profile fixed to their central values.} 

\begin{center}
\begin{tabular}{|c|c|c|}
\hline
OBSDIST$\_$DIM (m) & 70   \\ 
J$_{0Q} $ & 0.22  \\
$\sigma_n$ (J/m$^2$)& 0.08\\
R & 0.3 \\
L (g/cm$^2$) & 220 \\
\hline
\end{tabular}
\tablab{mytable1}
\end{center}
\end{table}

With the improved parmetrizations of the current profile as given in \secref{ParamCurr}
we  validate the performance of MGMR3D by fitting the radio footprint of showers simulated with CoREAS to that of MGMR3D. There is a range of parameters available in the framework of MGMR3D that can be tuned to achieve a good fit. We follow the approach where generic shower parameters, based on shower generality, are taken fixed, such as those given in \tabref{mytable1}, while others, in particular those describing the longitudinal profile of the shower ($\Xmax$, the shower maximum, and E, the shower energy) are fitted for each shower.

CoREAS simulations are performed on a star-shaped layout of antennas with the center on the shower axis and 8 arms. Each arm contains 20 antennas, with a spacing of 25~m in the shower plane. The radio pulses are filtered between 30 -- 80~ MHz. 

The results of each CoREAS simulation for the intensity $I$ for all antennas of the grid, is fitted with MGMR3D using a steepest descent algorithm treating $\Xmax$ and $E$ as free parameters. In these calculations, the core position is kept fixed to the center of the grid. In later applications to LOFAR data (\secref{lofardata}) the core position is also treated as a free parameter

\subsection{Single shower comparisons}

\begin{figure*}[t]
\begin{center}
\includegraphics[width=2\columnwidth]{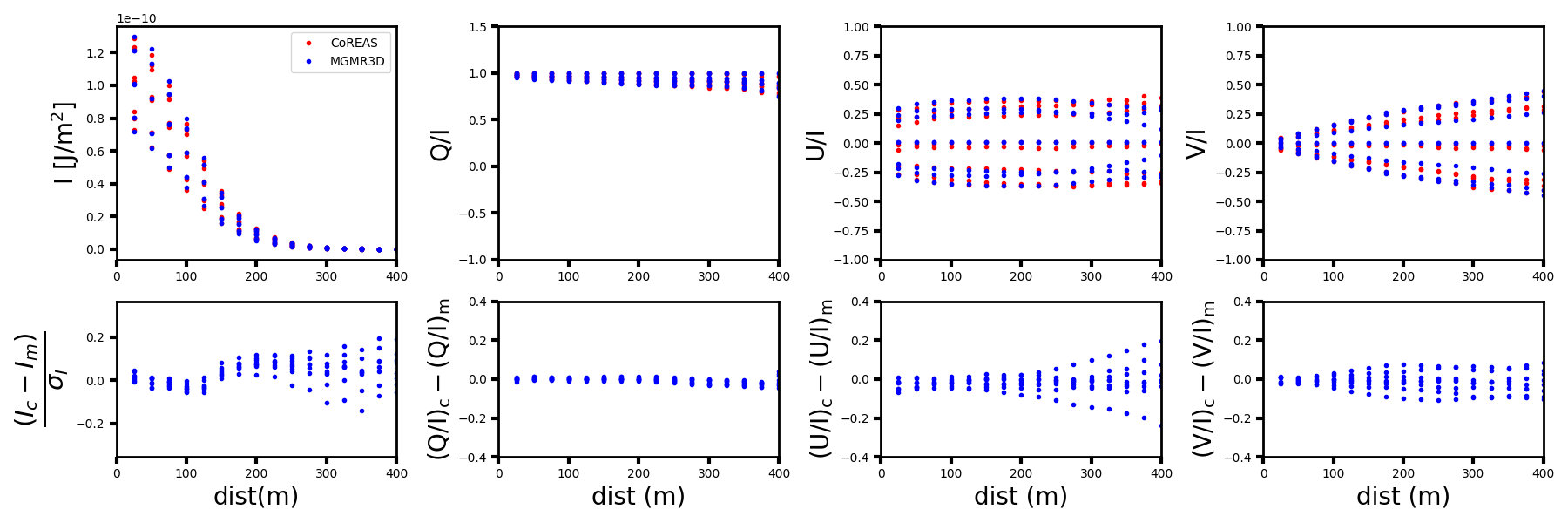}
\caption{Comparing Stokes parameters between  best fitting MGMR3D to CoREAS, for a 10$^{8.1}$ GeV shower with zenith angle=26$^\circ$ and $\Xmax=$~631 g/cm$^2$. The top panel in each plot shows the Stokes parameter as a function of antenna distance, the bottom panel shows the difference between CoREAS and MGMR3D. At each radial distance, there are eight data points corresponding to the antennas lying on different arms of the star-shaped layout.}
 \figlab{Stokes_lowzen}
\end{center}
\end{figure*}

\begin{figure*}[t]
\begin{center}
\includegraphics[width=2\columnwidth]{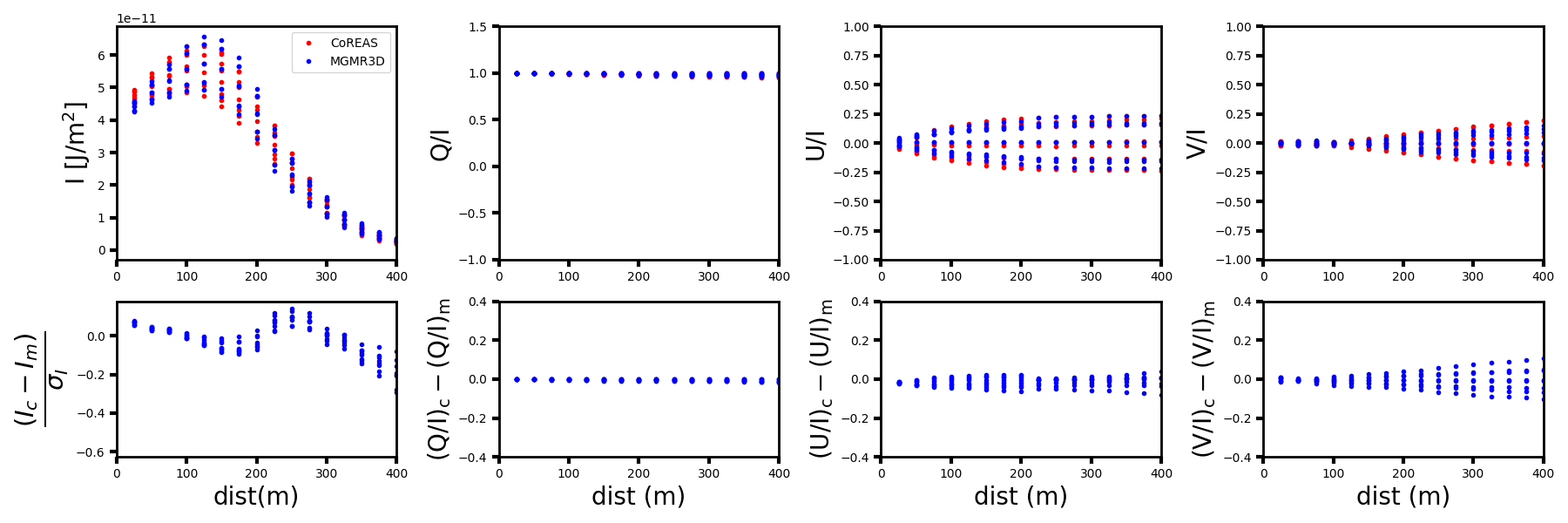}
\caption{Comparing Stokes parameters between  best fitting MGMR3D to CoREAS, for a  10$^{8.2}$ GeV shower with zenith angle=46$^\circ$ zenith with $\Xmax=$~630 g/cm$^2$. 
The top panel in each plot shows the Stokes parameter as a function of antenna distance, the bottom panel shows the difference between CoREAS and MGMR3D. At each radial distance, there are eight data points corresponding to the antennas lying on different arms of the star-shaped layout.}
\figlab{Stokes_highzen}
\end{center}
\end{figure*}

\begin{figure*}[t]
\begin{center}
\includegraphics[width=2\columnwidth]{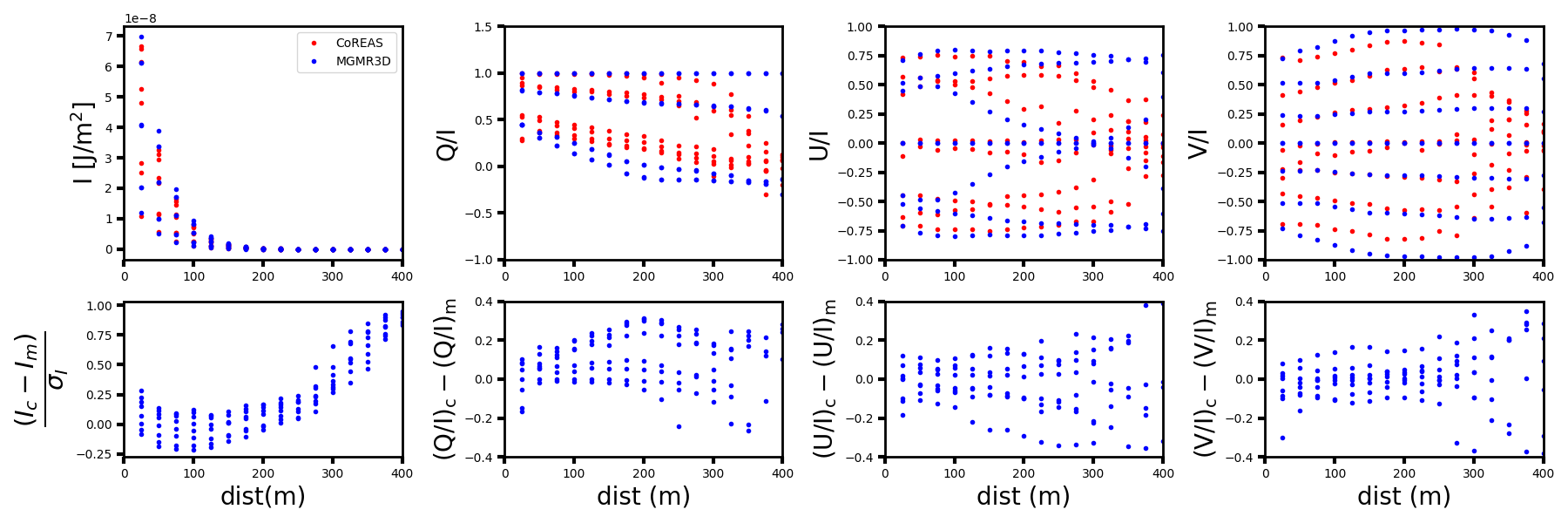}
\caption{Comparing Stokes parameters between  best fitting MGMR3D to CoREAS, for a shower with  very high $\Xmax=$~940 g/cm$^2$ with zenith angle=10$^\circ$ and shower energy $\approx$ 10$^9$ GeV, where the fit quality is poor.}
\figlab{Stokes_highxmax}
\end{center}
\end{figure*}


The different panels in \figref{Stokes_lowzen} and \figref{Stokes_highzen} show the Stokes parameters for two showers coming in at a 26$^\circ$ and 46$^\circ$ zenith respectively.  The top panels show the Stokes parameter as a function of antenna position for both MGMR3D and CoREAS and the bottom panels show the relative difference between the two models defined as $\Delta{I}= \frac{(I_c-I_m)}{\sigma_I}$. The realistic error model described  in \eqref{errormodel} is used. 
All the plots show a common feature that the magnitude of $\Delta{I}$ varies with antenna positions and has zero crossings.

The magnitudes of the Stokes parameters depend on the azimuthal orientations of the antennas with respect to the core. 
For example,  along the $\bf{v} \times \bf{B}$ direction there is full linear polarization resulting $Q/I=1$. It deviates from unity for other  directions,
due to a small contribution from the charge-excess emission. 
Similarly, the circular polarization,
expressed by $V/I$, is small and azimuth angle dependent.

The  Stokes parameters $U$ and $V$ for the two calculations are shown to agree well within 250 meters, while the differences  increase at larger distances. These differences  seem to point to an underestimate of the difference in emission heights between charge excess and transverse current radiation in MGMR3D.

\figref{Stokes_highxmax} shows an example of a shower with a very large  $\Xmax$ $\approx$ 950 g/cm$^2$ which results in a poor agreement between CoREAS and MGMR3D, such cases can be expected when the shower develops closer  to the ground. Further details for such cases are discussed in \secref{simu_fit}. 


In the rest of this work,  we concentrate on reconstructing the shower maximum using Stokes I. We restrict ourselves to $I$ as it is the Stokes parameter that can most accurately be measured experimentally, and we have also noted that adding other Stokes parameters does not lead to any significant improvement in the reconstruction of air shower parameters.

\subsection{Fitting the shower maximum} \seclab{simu_fit}

In this section,  we report the results of reconstructing $\Xmax$ with MGMR3D by fitting an ensemble of CoREAS showers. This CoREAS library was produced for  each detected shower in LOFAR, where at least 25 proton and 10 iron  showers are simulated with the same energy and arrival direction obtained from a preliminary reconstruction for this shower \cite{lofar}.

The radio footprints with MGMR3D are fitted to CoREAS with $\Xmax$ as a free parameter for each shower with arrival direction and energy same as CoREAS. As mentioned earlier, for CoREAS simulations the shower core positions are known, hence we do not fit the core positions. But for real data  core positions become important fit parameters while obtaining the  radio profile that best describes the data. This is discussed in detail in the next section.

\begin{figure}[t]

\includegraphics[width=1.\columnwidth]{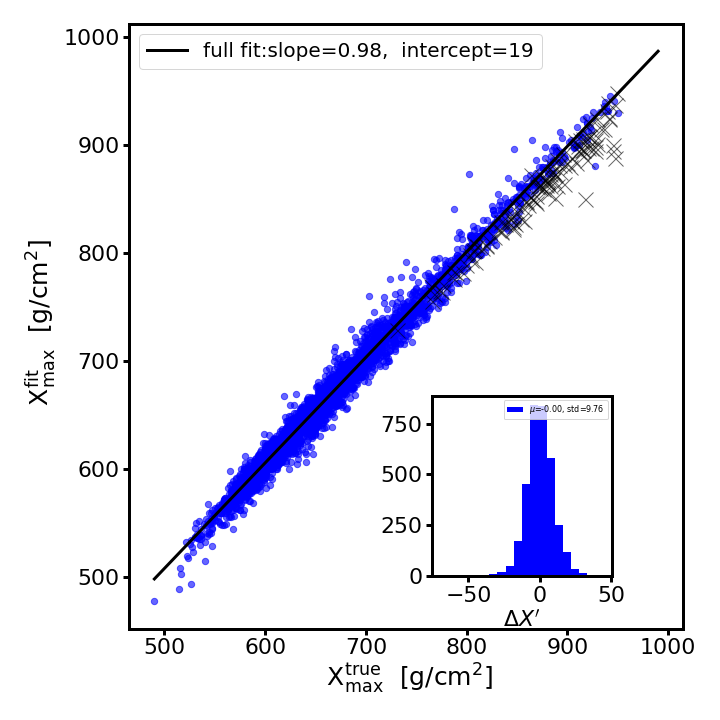}

\caption{Scatter plot of reconstructed $\Xmax$ with MGMR3D and CoREAS. The ensemble contains showers with different energy, zenith, and azimuth angles and for each shower at least 25 proton and 10 iron simulations are considered. A quality cut based on the distance from the core of the shower on the
ground to $\Xmax$ is applied. The black crosses represent showers excluded by the cut. The straight line is the best fitting line to the selected points passing the cut (blue). Distribution of the deviation of  $\Xmax^{\mathrm{fit}}$ from the fitting line, denoted by $\Delta X' $ is shown in the inset histogram. The resolution of the fit is 9.76 g/cm$^2$. }
\figlab{xmaxfit_coreas}
\end{figure}

We refer to the $\Xmax$ values obtained from CORSIKA as $\Xmax^{\mathrm{true}}$ and the reconstructed values as $\Xmax^{\mathrm{fit}}$.
The results are shown in \figref{xmaxfit_coreas}. This considers mixed primaries with proton and iron for various showers. The error calculated for the realistic noise model given in \eqref{errormodel} is used. We have applied a quality cut based
on the distance to $\Xmax$ from the ground. Details of this cut are explained in the following paragraphs. The black crosses are the points that are excluded by the cut.
A straight line is fit through the selected points, shown by the blue points. 
It is evident that there is a very strong correlation between the reconstructed $\Xmax$ and the CoREAS truth values. The slope and intercepts of the fit are 0.98 and 19 respectively. Distribution of the deviation of  $\Xmax^{\mathrm{fit}}$ from the fitting line, denoted by $\Delta X' $ is shown in the inset histogram of \figref{xmaxfit_coreas}. This shows a resolution of 9.76 g/cm$^2$.
It is also worth mentioning that we have studied the fits on proton and iron showers separately and  found no bias on primary particle type. The fit results are found to be almost identical, we have thus used combined showers for the rest of the analysis.

\begin{figure}[t]
\includegraphics[width=1.\columnwidth]{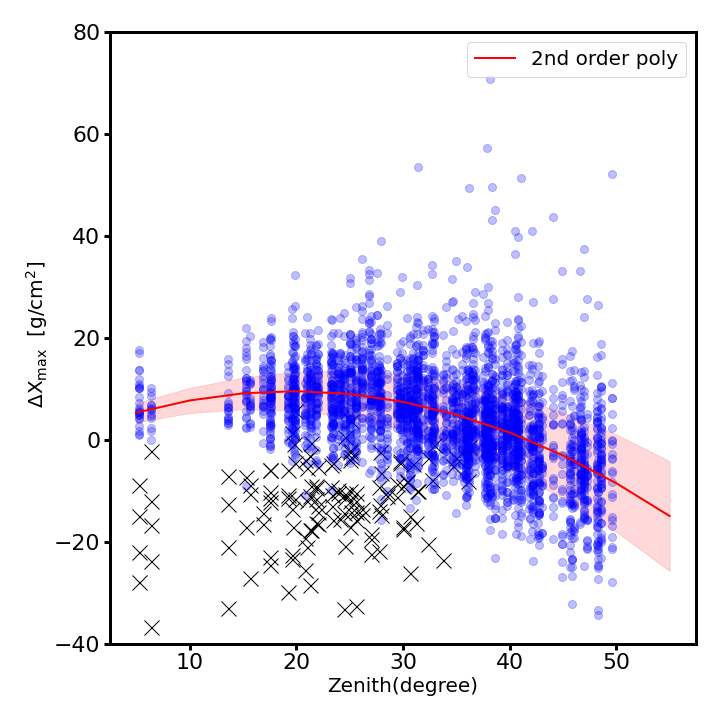}
\caption{Zenith angle distribution of $\Delta\Xmax$ of the showers shown in \figref{xmaxfit_coreas}. The black crosses are showers having distance to $\Xmax$ lower than 3 km and are excluded from the fit. The red line is a 2$^{nd}$ degree polynomial fit to the showers with the dashed line showing the 1-$\sigma$ deviation. The fit coefficients are shown in \tabref{mytable2}.}
\figlab{zenith_correction}
\end{figure}


The shift in $\Xmax$ from the true value is defined as
$\Delta\Xmax= {\Xmax}^{\mathrm{fit}} - {\Xmax}^{\mathrm{true}}$. For the majority of the showers, $\Delta\Xmax$ is independent of $\Xmax$ to the first order, as suggested by the near unity slope. 

However, we have found a  dependence on the shower zenith angle, as shown in \figref{zenith_correction}, which includes the same showers as in \figref{xmaxfit_coreas}. We see from the plot that  there are a handful of outliers, a few in the positive direction of $\Delta\Xmax$ and more in the negative direction.  The positive ones will be discussed in the next section. The negative outliers appear to be from showers that are developed closer to the ground. 
In order to obtain a clean parametrization to capture the relationship between $\Delta\Xmax$ and zenith, we have used a cut on the outliers.
These outliers are excluded based on a cut on distance from the core of the shower on the ground to $\Xmax$. We have chosen a conservative cut to accept showers with distance to $\Xmax$ $>$ 3 km in the fit that captures the trend between $\Delta\Xmax$ w.r.t. zenith, shown  by the red curve. The excluded points are shown in black crosses.  For  showers that are developed closer to the observer there are systematic differences between MGMR3D and CoREAS (also shown in an radio LDF example in \figref{Stokes_highxmax}), which could be attributed to the facts that for such showers more detailed parameters, like the dependence on the distance to the shower axis of the thickness and shape of the shower front, start to become important for the radio footprint,  leaving room for  more fine-tuning for specific showers with MGMR3D. Another important point is that, the general emission mechanism in MGMR3D involving coherence and farfield assumptions start to become less accurate when the emission is generated close to the antennas. However, for the majority of the showers the generic approximations hold and results with MGMR3D are in good agreement with COREAS.
  
The coefficients  of the fit from \figref{zenith_correction} are given in \tabref{mytable2}. This parametrization can be used as a correction factor to estimate the expected $\Xmax$ value from $\Xmax^{fit}$ in general and is used while fitting LOFAR data to MGMR3D in section \secref{lofardata}. 

\begin{table}[!ht]
\caption{ Polynomial fit of the form $\Delta\Xmax$ = p$_0$ $\theta^2$ + p$_1$ $\theta$ + p$_2$, where $\theta$ is the zenith of the shower in degree and  $\Delta\Xmax$ is the difference in $\Xmax$ fitted with MGMR3D from the CoREAS truth value.}
\begin{center}
\begin{tabular}{ |c|c|c| }
\hline
 p$_0$ &  p$_1$ & p$_2$ \\
 \hline
 -1.94 $\times$ 10$^{-1}$ & 0.769 & 1.56 \\
 \hline
 \end{tabular}
 \tablab{mytable2}
\end{center}
\end{table}

\subsection{Sensitivity to shower shape parameters-R and L}
It appears from \figref{zenith_correction} that there are a few showers,   where the $\Xmax$ fit from MGMR3D is  overestimated significantly from their CoREAS truth values that are not affected by the distance to $\Xmax$ cut described in the previous section. In this section, we take a closer look at some of these  cases.

\begin{figure}[t]
\includegraphics[width=1\columnwidth]{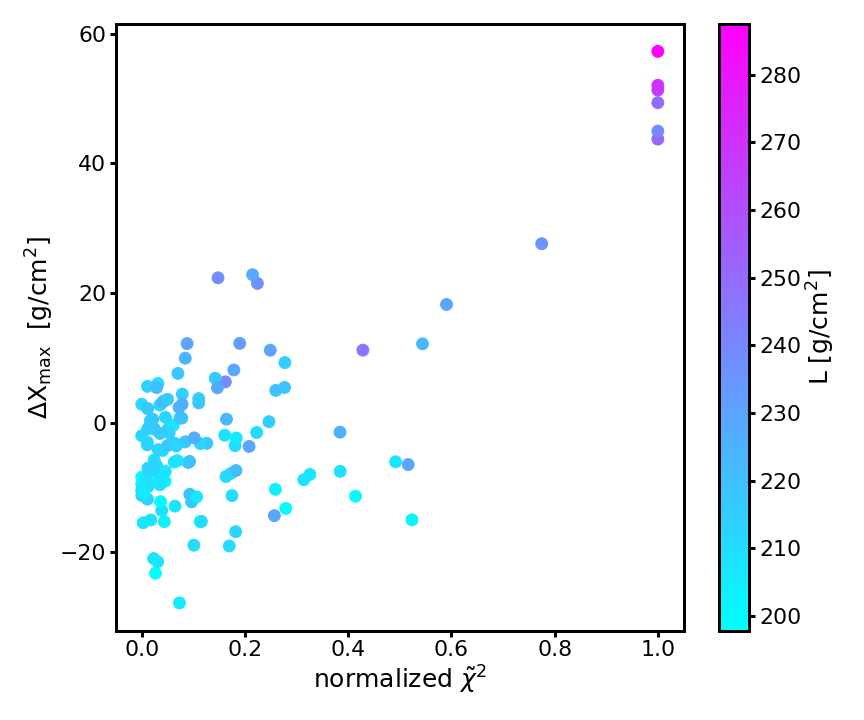}
\includegraphics[width=1\columnwidth]{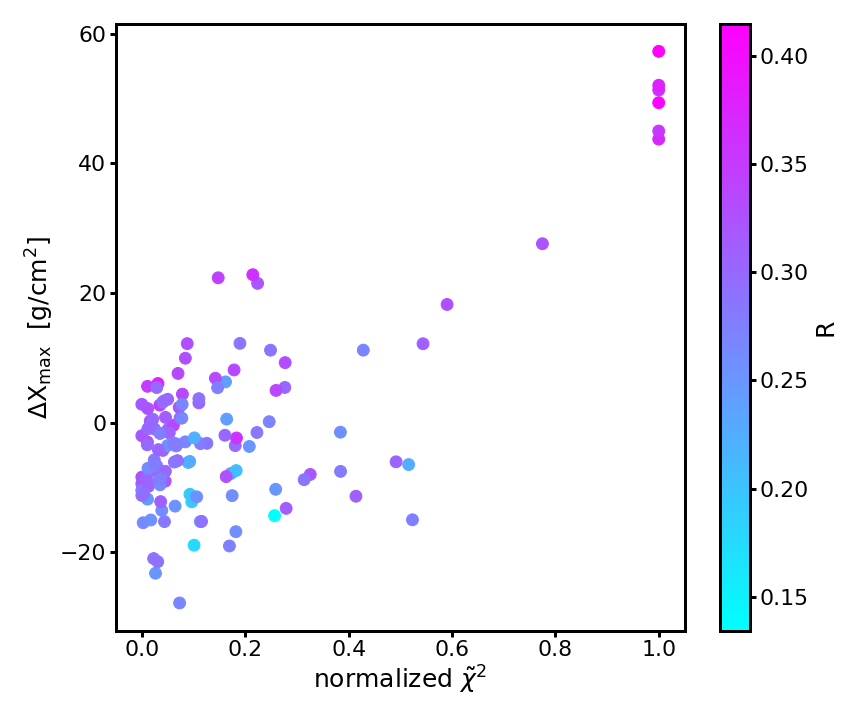}
\caption{ $\Delta\Xmax$ as function of $\tilde{\chi}^2$ with colorbars indicating true $L$ (top) and $R$ values (bottom) for an ensemble of proton simulations for a subset of showers. The $\tilde{\chi}^2$ values are normalised between 0 to 1 individually for each set of simulated showers. }
\figlab{LR_extreme}
\end{figure}

It is found that  these outliers have significantly larger $\tilde{\chi}^2$ values than the other CoREAS simulations for the same shower angle and energy. We have ruled out the possibility of non-convergence of the fit, by studying the $\tilde{\chi}^2$  surface for $\Xmax$, which showed a clear global minimum for all cases. While probing other reasons for such differences, we have found that these showers have longitudinal profiles that differ considerably from the rest of the ensemble. These differences are observed in terms of the shape parameters - $R$ and $L$ as described in \eqref{Def-Nc-GH}. It appears that for these outliers the true $R$ and $L$ values, obtained from fitting the CORSIKA longitudinal profiles, are quite extreme compared to their central values. A zoom of the subset of showers containing the outliers are shown in \figref{LR_extreme}, with their true $R$ and $L$ color coded. The trend demonstrates the correlation between the high $\Delta\Xmax$ with high $\tilde{\chi}^2$  (normalised between 0-1), and extreme $R$ and $L$. 

\begin{figure}[t]
\includegraphics[width=1.1\columnwidth]{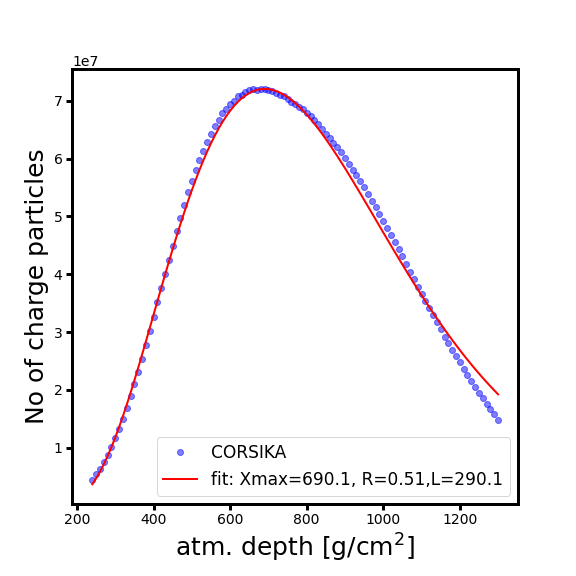}
\caption{  Longitudinal profile for a simulated shower. This is one of the extreme cases shown in \figref{LR_extreme}.  The dots are the CORSIKA values and the line is fit to the profile.}
\figlab{LR_property}
\end{figure}


The CORSIKA longitudinal profile for the extreme case is shown in \figref{LR_property}.
The shower shape is wider than usual and  this could indicate  the presence of an energetic secondary shower. 

In MGMR3D the $R$ and $L$ parameters are fixed to central values (see \tabref{mytable1})and we fit $\Xmax$ only, this can explain the large  shift in predicted $\Xmax$ which arises to compensate for the difference in longitudinal profile, however, the $\tilde{\chi}^2$ for these outliers still remains higher than the ensemble. This  example clearly shows two important results.  Firstly, the radio profiles are influenced by other parameters of the longitudinal profile than only $\Xmax$. Secondly,  MGMR3D is sensitive to these parameters. To extract all three parameters- $R$, $L$, and $\Xmax$, from  MGMR3D calculation requires more dedicated efforts and currently is beyond the scope of this paper. However, the outliers are only a small fraction of the total number of showers, and this would have only a small effect on the zenith based correction proposed in \tabref{mytable2}.


\section{Application to LOFAR data}\seclab{lofardata}


In this section we discuss various steps of  applying MGMR3D to experimental data and estimate $\Xmax$. We have used LOFAR cosmic ray data for this purpose.
Currently, LOFAR provides the highest precision for the determination of $\Xmax$ with the radio technique~\cite{xmax}. The dense core of LOFAR consists of 288 low-band dipole antennas within an area with a diameter of 320 meters, known as the Superterp. 
The radio emission from air showers in the frequency range 30 -- 80~MHz is recorded by the LOFAR low-band antennas \cite{LOFAR_haarlem}. An array of particle detectors, LORA, installed on the Superterp provides the trigger for the detection of the air showers ~\cite{lora}.

The usual $\Xmax$ reconstruction technique used at LOFAR is based on
the production of dedicated CoREAS simulation sets for each detected air
shower. The number of simulations needed to reconstruct the
shower maximum is optimized with CONEX\cite{conex}. A set of  CORSIKA simulations with proton and iron primaries is produced for each detected cosmic ray. 
The radio emission is simulated in a star-shaped pattern for antenna positions in the shower plane using CoREAS.
For each CoREAS simulation the value of $\Xmax$ as 
well as the $\chi^2$ is determined when fitting the core position to data. 
$\Xmax$ for a measured shower is then
reconstructed by fitting a parabola to the $\chi^2$ vs Monte
Carlo $\Xmax$ contour. The latest results on LOFAR cosmic ray analysis can be found in \cite{arthur_masscomp}. While such a Monte-Carlo based approach is precise, it is compute-intensive. Thus, fast alternatives such as MGMR3D are desired, where $\Xmax$ is reconstructed in a steepest descent optimization of the parametrized radio
profile to given data.
 
 The details of applying MGMR3D to data are as followed-  the quantity $P_{\rm data}$ or $P_{\rm mgmr3d}$, is calculated as the time integrated voltage squared over a 55~ns window centered around the pulse maximum, and is used as
the observable. The  error, $\sigma_P$, is estimated from the measurement of the noise level from data. This is the  same procedure as used in \cite{arthur_masscomp}. 
This implementation is different from the previous case of fitting only to simulations where the stokes parameters, integrated over the full trace,  were used as observables. 

The reduced $\chi^2$ to be minimized in MGMR3D is defined as
\beq
\chi ^2 =\frac{1}{N} \sum_{antennas} \left(\frac{P_{\rm data} -  P_{\rm mgmr3d} (x_{\rm core}, y_{\rm core}, \Xmax) }{\sigma_P}\right)^2 \;,
\eqlab{eq_norm}
\eeq
$\Xmax$ and the core positions ($x_\mathrm{core}$, $y_\mathrm{core}$) are the free parameters of the fit. The shower energy for the MGMR3D calculation is determined from the normalization constant, see \eqref{norm}.


In fitting to the data we have kept the longitudinal shape parameters $R$, and $L$ as well as the charge excess parameter J$_{0Q}$ fixed to the values given in \tabref{mytable1}. Including these parameters in the fit sometimes gave rise to a poor convergence without considerably improving the fit quality.

The core reconstruction from a parametrization of the radio LDF ~\cite{anna} is used as initial guesses for the core positions, the same as was also used in the CoREAS reconstruction method. In order to fit $\Xmax$, it is seen that starting from a small value  between 300-400~g/cm$^2$ leads to faster convergence.

\begin{figure}[th]
\includegraphics[width=1.\columnwidth]{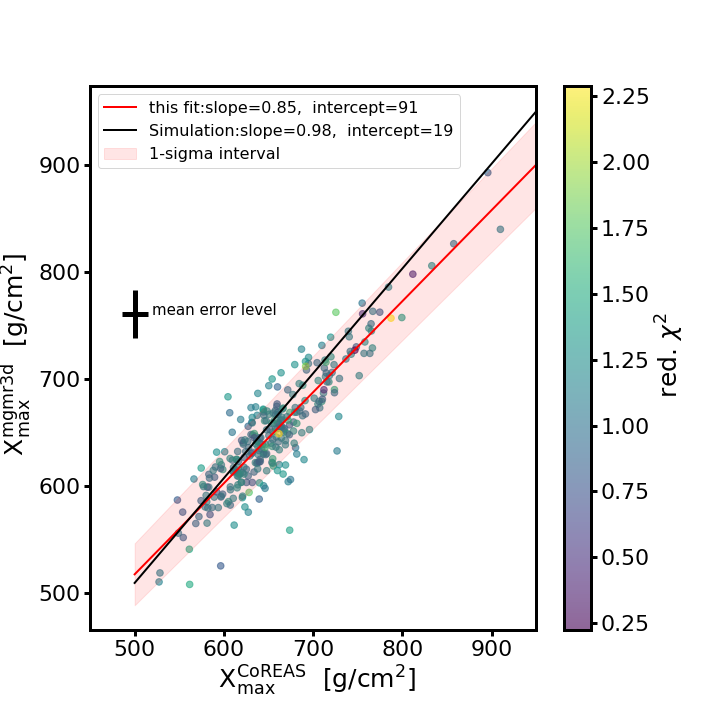}
\caption{Reconstructed $\Xmax$ with MGMR3D and LOFAR. The $\Xmax$ values reconstructed with MGMR3D are corrected with the zenith correction from \tabref{mytable2}. The mean error bar is shown by the black cross. The red line is the fit to the data points considering the vertical mean error in the fit. The shaded area is the 1-$\sigma$ interval of the fit. The black line corresponds to the prediction from simulation as discussed in \secref{simu_fit}. 
}
\figlab{xmaxfit1}
\end{figure}


The reconstructed  $\Xmax$ with MGMR3D are shown in comparison to the obtained $\Xmax$ using the LOFAR reconstruction technique in \figref{xmaxfit1}. The $\Xmax$ values  reconstructed with MGMR3D are corrected with the zenith correction formula described in \tabref{mytable2}. We have also implemented the distance to $\Xmax$ based quality cut  as described in \secref{simu_fit}. The red line is a linear fit to the data  with a slope of 0.85 and intercept 91. The  shaded area is the 1-$\sigma$ error on the fit. 
The black line is the prediction from simulations only, as discussed (cf. \figref{xmaxfit_coreas}). 

\begin{figure}[ht]
\includegraphics[width=1.1\columnwidth]{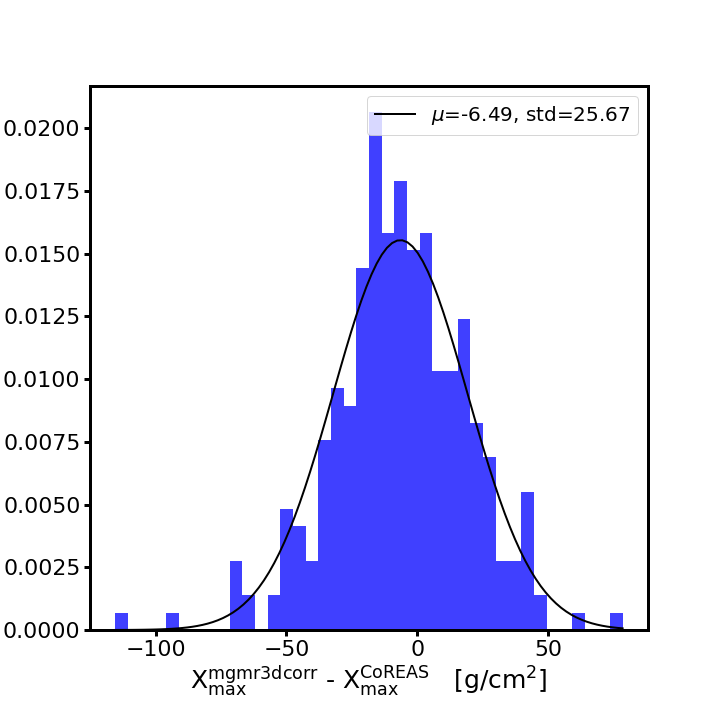}
\caption{The distribution of $\Xmax^{\rm{mgmr3d}}-\Xmax^{\rm{CoREAS}}$ is fitted by a Gaussian with $\mu$=-6.49~g/cm$^2$ and a standard deviation $\sigma_{tot}$=25.67~g/cm$^2$.}
\figlab{calc_err}
\end{figure}

From the comparison shown in \figref{xmaxfit1} an estimate can be obtained for the accuracy for $\Xmax^{\rm{mgmr3d}}$. 
The combined error on $\Xmax$ is calculated from the standard deviation  of the  gaussian fitted to the distribution of $\Xmax^{\rm{mgmr3d}}$ - $\Xmax^{\rm{CoREAS}}$ as shown in \figref{calc_err}.
Assuming the errors due to MGMR3D and CoREAS reconstruction are uncorrelated the total error
$\sigma_{tot}$ can be written as
\bea
\sigma^2_{tot} = \sigma^2_{\mathrm{coreas}} + \sigma^2_{\mathrm{mgmr3d}},
\eea
$\sigma_{\rm{coreas}}$ is obtained from the mean of the distribution of errors on $\Xmax$ reconstructed with CoREAS  for individual events, using a Monte-Carlo method \cite{xmax}.
With  $\sigma_{\rm{coreas}}= 14.5$~g/cm$^2$ we obtain $\sigma_{\rm{mgmr3d}}= 22.4$ g/cm$^2$. This value is used as the resolution of the $\Xmax$ reconstruction with MGMR3D from LOFAR data and shown in the  black cross in \figref{xmaxfit1}. Since for CoREAS the shower is given by a microscopic CORSIKA calculation,  it is possible to obtain the error on $\Xmax$ from the quality of the fit but for MGMR3D such a procedure is not possible. The reason is that in MGMR3D calculations, parameters entering in the longitudinal profile, can easily vary well outside the physical regime. 

An example of the radio profile of a reconstructed shower is shown in Appendix A for both CoREAS and MGMR3D.

\subsection{Reconstruction of shower core and energy:}


\begin{figure}[t]
\includegraphics[width=1.\columnwidth]{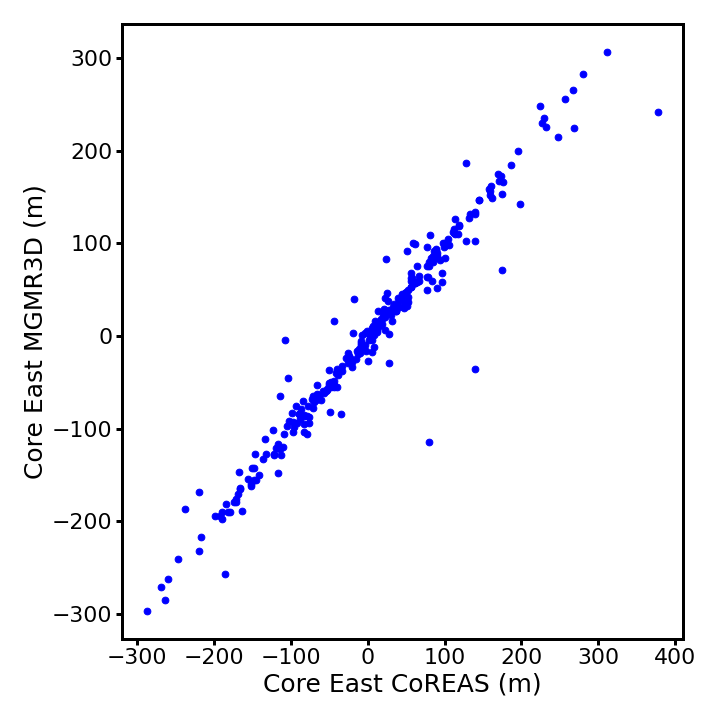}
\includegraphics[width=1.\columnwidth]{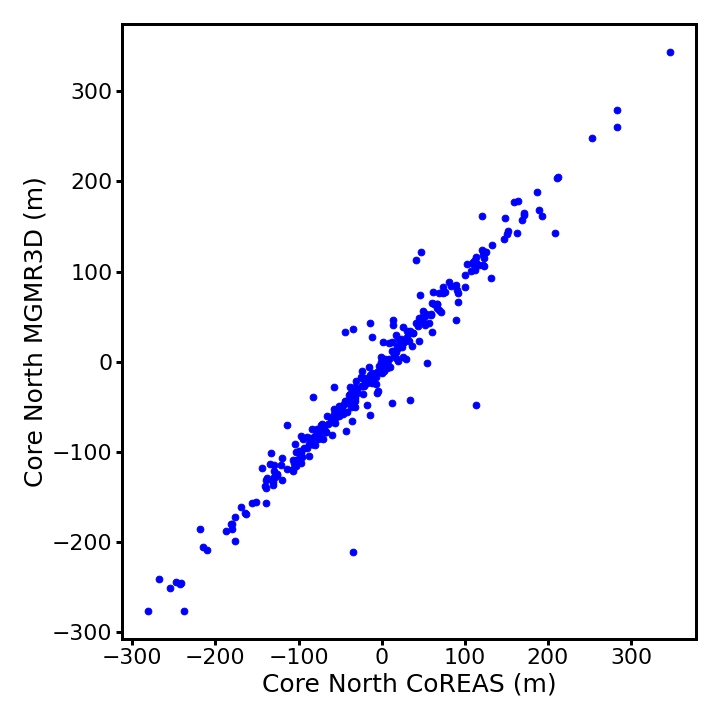}
\caption{Correlation in reconstructed cores for all showers  between CoREAS and MGMR3D for two directions .}
\figlab{deltaxmax_coreshift}
\end{figure}

In \figref{deltaxmax_coreshift} we show the correlation between the core positions reconstructed using MGMR3D and CoREAS reconstructed core positions. 
For the majority of the showers, the core positions show good agreement between COREAS and MGMR3D reconstructions. 
However, there are a few exceptions with large deviations between MGMR3D and CoREAS. This effect is not found to be correlated either with $\Delta \Xmax$ nor $\chi^2$. Some of these events are hard to reconstruct because the signal-to-noise ratio is relatively low, while others have a core that it is not well-contained by the LOFAR stations. In both cases, small differences between CoREAS and MGMR3D can have an impact that is larger than usual. 







\begin{figure}[t]
\includegraphics[width=1\columnwidth]{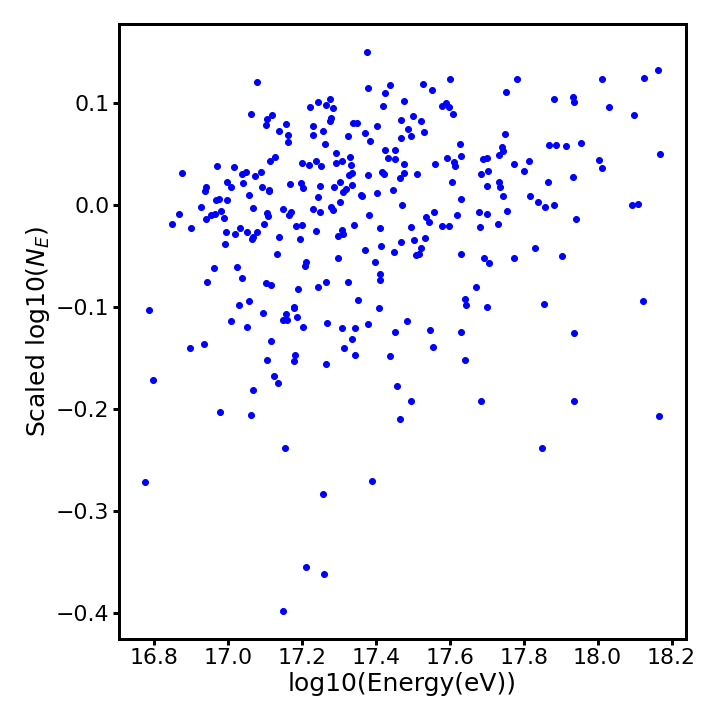}
\includegraphics[width=1\columnwidth]{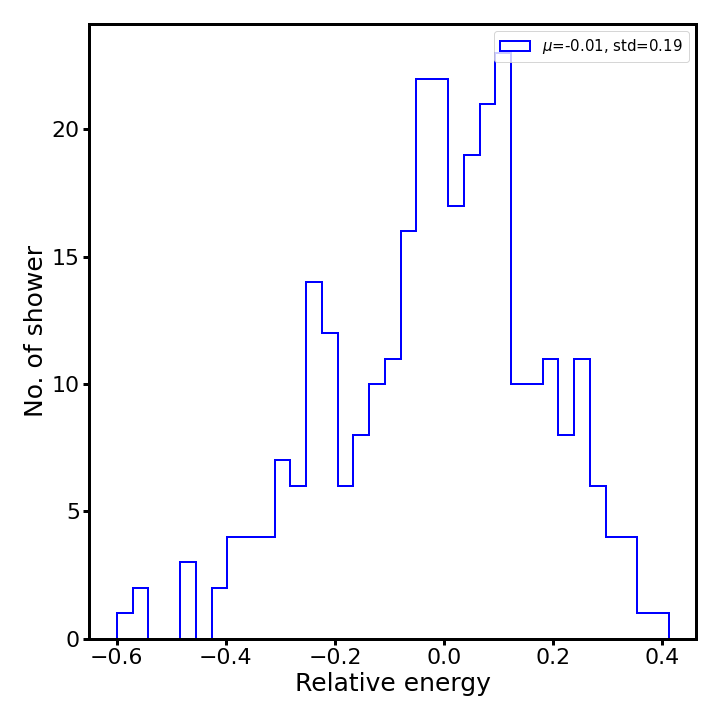}
\caption{(Top): The relative energies  $N_E=E_{\rm{MGMR3D}}/E_{\rm{CoREAS}}$ from the MGMR3D and CoREAS reconstructions of LOFAR data as a function of the CoREAS values, $E_{\rm{CoREAS}}$. (Bottom): The distribution of relative energy defined in \eqref{rel_en}. }
\figlab{E_LOFAR}
\end{figure}

In \figref{E_LOFAR}  the differences in cosmic ray energy reconstruction
between MGMR3D, using \eqref{norm}, and CoREAS are compared.
The top panel of \figref{E_LOFAR} shows that there is no clear correlation between the two. The bottom panel of the figure shows the relative difference, defined as,
\beq
2\, \frac{(E_{\mathrm{MGMR3D}}-E_{\mathrm{CoREAS}})} {(E_{\mathrm{MGMR3D}}+E_{\mathrm{CoREAS}})} 
\eqlab{rel_en}
\eeq
to make the differences more quantitative.  
This shows that there is no average offset between the two energy reconstructions. The spread of 19\% in the distribution is comparable to the LOFAR energy resolution of 14\% ~\cite{katie_energyscale}.  



\section{Summary and conclusions}
The MGMR3D code, which uses  an analytic parameterization of the  plasma cloud,  provides a  promising alternative to obtain the longitudinal structure of an air shower that best reproduces the measured radio footprint through minimization. It is computationally orders of magnitude faster than its microscopic counterparts that are customarily used for analyzing radio emission from cosmic rays. 
We  have reported on a detailed comparison for a large ensemble of showers simulated with CoREAS and MGMR3D. This resulted in an optimized parameterization inside MGMR3D, in particular concerning the drift velocity,  the charge excess, and the radial structure.  With the optimized parametrization a strong agreement with microscopic CoREAS-simulations were obtained for the lateral distribution functions for radio emission with a relative difference in intensity up to 10$\%$. 

As a follow-up  step we have shown that MGMR3D can be used in a chi-square fit procedure to extract the shower maximum $\Xmax$ for a large ensemble of showers simulated by CoREAS. The results show a very good agreement with a small systematic zenith-angle dependency, which is upto 6-8 g/cm$^2$ for zenith angles not exceeding 50 degrees. We introduce a correction formula to compensate for this. However, MGMR3D is yet not fully optimized for highly inclined showers with zenith above 65 degrees. This is a prospect for a future effort and would be useful for simulation studies for experiments such as GRAND(The Giant Radio Array for Neutrino Detection) designed for detecting highly inclined air showers. 

We have also found that MGMR3D is sensitive to the effects of additional parameters corresponding to the shape of the longitudinal shower profile on the radio footprint- namely R and L. These parameters have the potential to provide further insight in mass composition, constraining hadronic model, as well as astrophysical interpretation of cosmic ray sources, in addition to $\Xmax$~\cite{Buitink:2021pkz}. Probing these subtle parameter spaces require extremely dense antenna layouts such as The Square Kilometer Array(SKA)~\cite{Corstanje:2023uyg}, and the required simulations also multiply by many folds, which is exhaustive for present compute-intensive Monte Carlo frameworks. MGMR3D, thus, opens up the novel opportunity of making such multi-parameter study plausible by producing large simulation sets with very little compute resources. A detailed study along these lines will be investigated in a follow up work.

As a final proof of the proposed procedure we  have used MGMR3D to extract $\Xmax$ from LOFAR data that have been used in earlier studies. 
An average $\Xmax$ resolution of 22~g/cm$^2$ is found which is competitive to the average resolution of 14.5g/cm$^2$ obtained using the CoREAS based method. It shows  that,  the latest version of MGMR3D, for specific geometries discussed in this paper, can be used as a fast and efficient tool to reconstruct shower parameters, and for high-precision studies, it can be combined with Monte Carlo simulations as a preliminary estimator to help reduce the required simulation landscape and expedite the analysis.

\appendix
\section {Example from LOFAR data}

A comparison of the radio profiles between CoREAS and MGMR3D for one measured shower is shown in \figref{ldf_fit1} and the corresponding reconstructed parameters are in \tabref{table_reco}.

\begin{table}[ht]
\begin{tabular} {c c c c c}

& No& $\Xmax$ & red. $\chi^2$& core pos. \\
&  & [g/cm$^2$]& red. $\chi^2$& [E,N]~(m) \\
\hline
 CoREAS & 1 & 679.3 & 1.15 &[102.1, -42.0] \\
 MGMR3D & 1 & 675.2 & 1.25 & [98.6 -55.6] \\
\end{tabular}
\caption{Table parameters for showers shown in \figref{ldf_fit1}. The core position is given as [east,north](m) with respect to the center of the Superterp.}
\tablab{table_reco}
\end{table}

\begin{figure}
\includegraphics[width=1.\columnwidth]{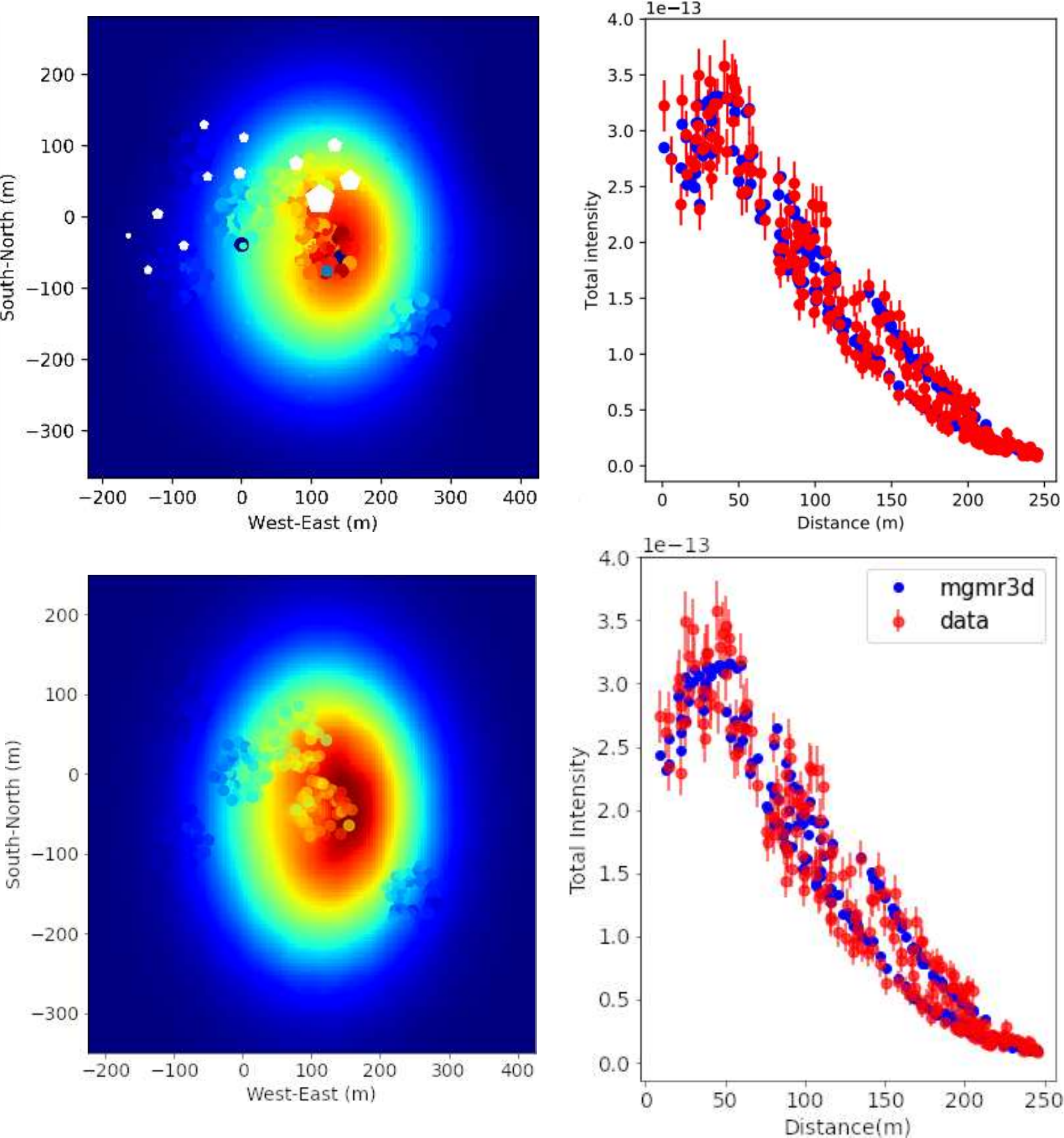}
\caption{ Example of a measured shower fitted with CoREAS (top panel) and MGMR3D(bottom). Top panel: (left) antenna configuration on the ground with color coded intensity with the best-fitting simulated footprint with CoREAS on the background.  The one-dimensional LDF intensity profile for CoREAS (right). This figure is adapted from LOFAR reconstruction library catalogs, and also features the particle detectors shown by the white hexagons. Bottom panel: (left and right) similar to the panel above but here simulations are done with MGMR3D. For reconstructed shower parameters see \tabref{table_reco}.  }
\figlab{ldf_fit1}
\end{figure}

\section{Programming details}
The latest version of the program can be downloaded from \cite{olaf_scholten_2023_7698097}. This version contains the improved parametrizations, realistic errormodel discussed in this paper, as well as the functionality to include antenna response functions, relevant for the application to measured data.    
\section{Acknowledgement}
P.~Mitra acknowledges financing by the Polish National Agency for Academic Exchange within Polish Returns Program no. PPN/PPO/2020/1/00024/U/00001. This research is also funded by Vietnam National Foundation for Science and Technology Development (NAFOSTED) under Grant No. 103.01-2019.378. BMH is funded by ERC Grant agreement No. 101041097. N.Karastathis acknowledges funding by the Deutsche Forschungsgemeinschaft (DFG, German Research Foundation) – Projektnummer 445154105. ST acknowledges funding from the Khalifa University Startup grant, project code 8474000237-FSU-2020-13. LOFAR, the Low Frequency Array designed and constructed by ASTRON, has facilities in several countries, that are owned by various parties (each with their own funding sources), and that are collectively operated by the International LOFAR Telescope foundation under a joint scientific policy.
\clearpage
\newpage
\bibliography{LOFARbib_new}

\begin{thebibliography}{33}
\expandafter\ifx\csname natexlab\endcsname\relax\def\natexlab#1{#1}\fi
\expandafter\ifx\csname bibnamefont\endcsname\relax
  \def\bibnamefont#1{#1}\fi
\expandafter\ifx\csname bibfnamefont\endcsname\relax
  \def\bibfnamefont#1{#1}\fi
\expandafter\ifx\csname citenamefont\endcsname\relax
  \def\citenamefont#1{#1}\fi
\expandafter\ifx\csname url\endcsname\relax
  \def\url#1{\texttt{#1}}\fi
\expandafter\ifx\csname urlprefix\endcsname\relax\def\urlprefix{URL }\fi
\providecommand{\bibinfo}[2]{#2}
\providecommand{\eprint}[2][]{\url{#2}}

\bibitem[{\citenamefont{Scholten et~al.}(2008)\citenamefont{Scholten, Werner,
  and Rusydi}}]{olaf_geosync}
\bibinfo{author}{\bibfnamefont{O.}~\bibnamefont{Scholten}},
  \bibinfo{author}{\bibfnamefont{K.}~\bibnamefont{Werner}}, \bibnamefont{and}
  \bibinfo{author}{\bibfnamefont{F.}~\bibnamefont{Rusydi}},
  \bibinfo{journal}{Astropart. Phys.} \textbf{\bibinfo{volume}{29}}
  (\bibinfo{year}{2008}).

\bibitem[{\citenamefont{Askaryan}(1962)}]{askaryan1}
\bibinfo{author}{\bibfnamefont{G.}~\bibnamefont{Askaryan}},
  \bibinfo{journal}{Soviet Phys. JETP} \textbf{\bibinfo{volume}{14}}
  (\bibinfo{year}{1962}).

\bibitem[{\citenamefont{Askaryan}(1965)}]{askaryan2}
\bibinfo{author}{\bibfnamefont{G.}~\bibnamefont{Askaryan}},
  \bibinfo{journal}{Soviet Phys. JETP} \textbf{\bibinfo{volume}{21}}
  (\bibinfo{year}{1965}).

\bibitem[{\citenamefont{Buitink et~al.}(2016)}]{nature}
\bibinfo{author}{\bibfnamefont{S.}~\bibnamefont{Buitink}} \bibnamefont{et~al.},
  \bibinfo{journal}{Nature} \textbf{\bibinfo{volume}{531}}, \bibinfo{pages}{70}
  (\bibinfo{year}{2016}), \eprint{1603.01594}.

\bibitem[{\citenamefont{Alvarez-Muniz et~al.}(2012)\citenamefont{Alvarez-Muniz,
  Carvalho, and Zas}}]{zharies}
\bibinfo{author}{\bibfnamefont{J.}~\bibnamefont{Alvarez-Muniz}},
  \bibinfo{author}{\bibfnamefont{W.~R.} \bibnamefont{Carvalho},
  \bibfnamefont{Jr.}}, \bibnamefont{and}
  \bibinfo{author}{\bibfnamefont{E.}~\bibnamefont{Zas}},
  \bibinfo{journal}{Astropart. Phys.} \textbf{\bibinfo{volume}{35}},
  \bibinfo{pages}{325} (\bibinfo{year}{2012}), \eprint{1107.1189}.

\bibitem[{\citenamefont{Huege et~al.}(2013)\citenamefont{Huege, Ludwig, and
  James}}]{Huege:2013vt}
\bibinfo{author}{\bibfnamefont{T.}~\bibnamefont{Huege}},
  \bibinfo{author}{\bibfnamefont{M.}~\bibnamefont{Ludwig}}, \bibnamefont{and}
  \bibinfo{author}{\bibfnamefont{C.~W.} \bibnamefont{James}},
  \bibinfo{journal}{AIP Conf. Proc.} \textbf{\bibinfo{volume}{1535}},
  \bibinfo{pages}{128} (\bibinfo{year}{2013}), \eprint{1301.2132}.

\bibitem[{\citenamefont{de~Vries et~al.}(2013)\citenamefont{de~Vries, Scholten,
  and Werner}}]{krijn_cherenkov}
\bibinfo{author}{\bibfnamefont{K.~D.} \bibnamefont{de~Vries}},
  \bibinfo{author}{\bibfnamefont{O.}~\bibnamefont{Scholten}}, \bibnamefont{and}
  \bibinfo{author}{\bibfnamefont{K.}~\bibnamefont{Werner}},
  \bibinfo{journal}{Astropart. Phys.} \textbf{\bibinfo{volume}{45}},
  \bibinfo{pages}{23} (\bibinfo{year}{2013}), \eprint{1304.1321}.

\bibitem[{\citenamefont{{ K. D. de Vries, O. Scholten }}(2013)}]{krijn_eva}
\bibinfo{author}{\bibnamefont{{ K. D. de Vries, O. Scholten }}},
  \bibinfo{journal}{AIP Conference Proceedings 1535, 133 (2013);}
  \textbf{\bibinfo{volume}{1535}}, \bibinfo{pages}{133} (\bibinfo{year}{2013}).

\bibitem[{\citenamefont{Scholten et~al.}(2018)\citenamefont{Scholten, Trinh,
  de~Vries, and Hare}}]{mgmr3d}
\bibinfo{author}{\bibfnamefont{O.}~\bibnamefont{Scholten}},
  \bibinfo{author}{\bibfnamefont{T.~N.~G.} \bibnamefont{Trinh}},
  \bibinfo{author}{\bibfnamefont{K.~D.} \bibnamefont{de~Vries}},
  \bibnamefont{and} \bibinfo{author}{\bibfnamefont{B.~M.} \bibnamefont{Hare}},
  \bibinfo{journal}{Phys. Rev. D} \textbf{\bibinfo{volume}{97}},
  \bibinfo{pages}{023005} (\bibinfo{year}{2018}),
  \urlprefix\url{https://link.aps.org/doi/10.1103/PhysRevD.97.023005}.

\bibitem[{\citenamefont{Jackson}(1999)}]{jacksonbook}
\bibinfo{author}{\bibfnamefont{J.~D.} \bibnamefont{Jackson}},
  \emph{\bibinfo{title}{Classical electrodynamics}}
  (\bibinfo{publisher}{Wiley}, \bibinfo{address}{New York, {NY}},
  \bibinfo{year}{1999}), \bibinfo{edition}{3rd} ed., ISBN
  \bibinfo{isbn}{9780471309321},
  \urlprefix\url{http://cdsweb.cern.ch/record/490457}.

\bibitem[{\citenamefont{Butler}(2020)}]{david_thesis}
\bibinfo{author}{\bibfnamefont{D.~A.} \bibnamefont{Butler}}, Ph.D. thesis,
  \bibinfo{school}{Karlsruher Institut für Technologie (KIT)}
  (\bibinfo{year}{2020}), \bibinfo{note}{51.03.04; LK 01}.

\bibitem[{\citenamefont{Zilles et~al.}(2020)}]{anna_radiomorph}
\bibinfo{author}{\bibfnamefont{A.}~\bibnamefont{Zilles}} \bibnamefont{et~al.},
  \bibinfo{journal}{Astroparticle Physics} \textbf{\bibinfo{volume}{114}},
  \bibinfo{pages}{10 } (\bibinfo{year}{2020}), ISSN \bibinfo{issn}{0927-6505},
  \urlprefix\url{http://www.sciencedirect.com/science/article/pii/S0927650518302883}.

\bibitem[{\citenamefont{Trinh et~al.}(2022)\citenamefont{Trinh, Scholten,
  Buitink, de~Vries, Mitra, Phong~Nguyen, and Si}}]{Trinh:2022}
\bibinfo{author}{\bibfnamefont{T.~N.~G.} \bibnamefont{Trinh}},
  \bibinfo{author}{\bibfnamefont{O.}~\bibnamefont{Scholten}},
  \bibinfo{author}{\bibfnamefont{S.}~\bibnamefont{Buitink}},
  \bibinfo{author}{\bibfnamefont{K.~D.} \bibnamefont{de~Vries}},
  \bibinfo{author}{\bibfnamefont{P.}~\bibnamefont{Mitra}},
  \bibinfo{author}{\bibfnamefont{T.}~\bibnamefont{Phong~Nguyen}},
  \bibnamefont{and} \bibinfo{author}{\bibfnamefont{D.~T.} \bibnamefont{Si}},
  \bibinfo{journal}{Phys. Rev. D} \textbf{\bibinfo{volume}{105}},
  \bibinfo{pages}{063027} (\bibinfo{year}{2022}),
  \urlprefix\url{https://link.aps.org/doi/10.1103/PhysRevD.105.063027}.

\bibitem[{\citenamefont{Kamata and Nishimura}(1958)}]{nkg}
\bibinfo{author}{\bibfnamefont{K.}~\bibnamefont{Kamata}} \bibnamefont{and}
  \bibinfo{author}{\bibfnamefont{J.}~\bibnamefont{Nishimura}},
  \bibinfo{journal}{Progress of Theoretical Physics Supplement}
  \textbf{\bibinfo{volume}{6}}, \bibinfo{pages}{93} (\bibinfo{year}{1958}),
  ISSN \bibinfo{issn}{0375-9687},
  \eprint{https://academic.oup.com/ptps/article-pdf/doi/10.1143/PTPS.6.93/5270594/6-93.pdf},
  \urlprefix\url{https://doi.org/10.1143/PTPS.6.93}.

\bibitem[{\citenamefont{Trinh et~al.}(2016)}]{gia1}
\bibinfo{author}{\bibfnamefont{T.~N.~G.} \bibnamefont{Trinh}}
  \bibnamefont{et~al.}, \bibinfo{journal}{Phys. Rev. D}
  \textbf{\bibinfo{volume}{93}}, \bibinfo{pages}{023003}
  (\bibinfo{year}{2016}),
  \urlprefix\url{https://link.aps.org/doi/10.1103/PhysRevD.93.023003}.

\bibitem[{\citenamefont{{Gaisser} and {Hillas}}(1977)}]{Gaisser:1977}
\bibinfo{author}{\bibfnamefont{T.~K.} \bibnamefont{{Gaisser}}}
  \bibnamefont{and} \bibinfo{author}{\bibfnamefont{A.~M.}
  \bibnamefont{{Hillas}}}, in \emph{\bibinfo{booktitle}{International Cosmic
  Ray Conference}} (\bibinfo{year}{1977}), vol.~\bibinfo{volume}{8} of
  \emph{\bibinfo{series}{International Cosmic Ray Conference}}, p.
  \bibinfo{pages}{353}.

\bibitem[{\citenamefont{Andringa et~al.}(2010)\citenamefont{Andringa, Concei{\c
  c}{\~a}o, and Pimenta}}]{rl_theory}
\bibinfo{author}{\bibfnamefont{S.}~\bibnamefont{Andringa}},
  \bibinfo{author}{\bibfnamefont{R.}~\bibnamefont{Concei{\c c}{\~a}o}},
  \bibnamefont{and} \bibinfo{author}{\bibfnamefont{M.}~\bibnamefont{Pimenta}},
  \bibinfo{journal}{{Astroparticle Physics}} \textbf{\bibinfo{volume}{34}},
  \bibinfo{pages}{360} (\bibinfo{year}{2010}),
  \urlprefix\url{https://hal.archives-ouvertes.fr/hal-00710483}.

\bibitem[{\citenamefont{Concei{\c{c}}{\~{a}}o
  et~al.}(2015)\citenamefont{Concei{\c{c}}{\~{a}}o, Andringa, Diogo, and
  Pimenta}}]{shape_RL}
\bibinfo{author}{\bibfnamefont{R.}~\bibnamefont{Concei{\c{c}}{\~{a}}o}},
  \bibinfo{author}{\bibfnamefont{S.}~\bibnamefont{Andringa}},
  \bibinfo{author}{\bibfnamefont{F.}~\bibnamefont{Diogo}}, \bibnamefont{and}
  \bibinfo{author}{\bibfnamefont{M.}~\bibnamefont{Pimenta}},
  \bibinfo{journal}{Journal of Physics: Conference Series}
  \textbf{\bibinfo{volume}{632}}, \bibinfo{pages}{012087}
  (\bibinfo{year}{2015}),
  \urlprefix\url{https://doi.org/10.1088/1742-6596/632/1/012087}.

\bibitem[{\citenamefont{Mitra}(2021)}]{Mitra:PhD}
\bibinfo{author}{\bibfnamefont{P.}~\bibnamefont{Mitra}}, Ph.D. thesis,
  \bibinfo{school}{Vrije Universiteit Brussel} (\bibinfo{year}{2021}),
  \urlprefix\url{https://www.vub.be/sites/vub/files/template_phd_pragati_mitra_-_eng.pdf}.

\bibitem[{\citenamefont{Schellart et~al.}(2013{\natexlab{a}})}]{pim_radio}
\bibinfo{author}{\bibfnamefont{P.}~\bibnamefont{Schellart}}
  \bibnamefont{et~al.}, \bibinfo{journal}{A\&A} \textbf{\bibinfo{volume}{560}},
  \bibinfo{pages}{A98} (\bibinfo{year}{2013}{\natexlab{a}}),
  \urlprefix\url{https://doi.org/10.1051/0004-6361/201322683}.

\bibitem[{\citenamefont{Scholten et~al.}(2016)}]{olaf_circpol}
\bibinfo{author}{\bibfnamefont{O.}~\bibnamefont{Scholten}}
  \bibnamefont{et~al.}, \bibinfo{journal}{Phys. Rev. D}
  \textbf{\bibinfo{volume}{94}}, \bibinfo{pages}{103010}
  (\bibinfo{year}{2016}),
  \urlprefix\url{https://link.aps.org/doi/10.1103/PhysRevD.94.103010}.

\bibitem[{\citenamefont{Dennis et~al.}(1981)\citenamefont{Dennis, Gay, and
  Welsch}}]{nl2sol}
\bibinfo{author}{\bibfnamefont{J.}~\bibnamefont{Dennis}},
  \bibinfo{author}{\bibfnamefont{D.}~\bibnamefont{Gay}}, \bibnamefont{and}
  \bibinfo{author}{\bibfnamefont{R.}~\bibnamefont{Welsch}},
  \bibinfo{journal}{ACM Trans. Math. Softw.} \textbf{\bibinfo{volume}{7}},
  \bibinfo{pages}{367} (\bibinfo{year}{1981}), ISSN \bibinfo{issn}{0098 --
  3500}.

\bibitem[{\citenamefont{Schellart et~al.}(2013{\natexlab{b}})}]{lofar}
\bibinfo{author}{\bibfnamefont{P.}~\bibnamefont{Schellart}}
  \bibnamefont{et~al.}, \bibinfo{journal}{Astronomy and Astrophysics}
  \textbf{\bibinfo{volume}{560}} (\bibinfo{year}{2013}{\natexlab{b}}).

\bibitem[{\citenamefont{Buitink et~al.}(2014)}]{xmax}
\bibinfo{author}{\bibfnamefont{S.}~\bibnamefont{Buitink}} \bibnamefont{et~al.},
  \bibinfo{journal}{Phys. Rev. D} \textbf{\bibinfo{volume}{90}}
  (\bibinfo{year}{2014}).

\bibitem[{\citenamefont{van Haarlem et~al.}(2013)}]{LOFAR_haarlem}
\bibinfo{author}{\bibfnamefont{M.~P.} \bibnamefont{van Haarlem}}
  \bibnamefont{et~al.}, \bibinfo{journal}{Astronomy and Astrophysics}
  \textbf{\bibinfo{volume}{556}}, \bibinfo{pages}{56} (\bibinfo{year}{2013}).

\bibitem[{\citenamefont{Thoudam et~al.}(2014)}]{lora}
\bibinfo{author}{\bibfnamefont{S.}~\bibnamefont{Thoudam}} \bibnamefont{et~al.},
  \bibinfo{journal}{Nucl.Instrum.Meth} \textbf{\bibinfo{volume}{A767}},
  \bibinfo{pages}{339} (\bibinfo{year}{2014}).

\bibitem[{\citenamefont{Bergmann et~al.}(2007)}]{conex}
\bibinfo{author}{\bibfnamefont{T.}~\bibnamefont{Bergmann}}
  \bibnamefont{et~al.}, \bibinfo{journal}{Astropart. Phys.}
  \textbf{\bibinfo{volume}{26}}, \bibinfo{pages}{420} (\bibinfo{year}{2007}),
  \eprint{astro-ph/0606564}.

\bibitem[{\citenamefont{Corstanje et~al.}(2021)}]{arthur_masscomp}
\bibinfo{author}{\bibfnamefont{A.}~\bibnamefont{Corstanje}}
  \bibnamefont{et~al.}, \bibinfo{journal}{arXiv:2103.12549}
  (\bibinfo{year}{2021}).

\bibitem[{\citenamefont{Nelles et~al.}(2015)}]{anna}
\bibinfo{author}{\bibfnamefont{A.}~\bibnamefont{Nelles}} \bibnamefont{et~al.},
  \bibinfo{journal}{Astropart. Phys.} \textbf{\bibinfo{volume}{65}},
  \bibinfo{pages}{11} (\bibinfo{year}{2015}), \eprint{1411.6865}.

\bibitem[{\citenamefont{Mulrey et~al.}(2020)}]{katie_energyscale}
\bibinfo{author}{\bibfnamefont{K.}~\bibnamefont{Mulrey}} \bibnamefont{et~al.},
  \bibinfo{journal}{Journal of Cosmology and Astroparticle Physics}
  \textbf{\bibinfo{volume}{2020}}, \bibinfo{pages}{017–017}
  (\bibinfo{year}{2020}), ISSN \bibinfo{issn}{1475-7516},
  \urlprefix\url{http://dx.doi.org/10.1088/1475-7516/2020/11/017}.

\bibitem[{\citenamefont{Buitink et~al.}(2021)}]{Buitink:2021pkz}
\bibinfo{author}{\bibfnamefont{S.}~\bibnamefont{Buitink}} \bibnamefont{et~al.},
  \bibinfo{journal}{PoS} \textbf{\bibinfo{volume}{ICRC2021}},
  \bibinfo{pages}{415} (\bibinfo{year}{2021}).

\bibitem[{\citenamefont{Corstanje et~al.}(2023)}]{Corstanje:2023uyg}
\bibinfo{author}{\bibfnamefont{A.}~\bibnamefont{Corstanje}}
  \bibnamefont{et~al.} (\bibinfo{year}{2023}), \eprint{2303.09249}.

\bibitem[{\citenamefont{Scholten}(2023)}]{olaf_scholten_2023_7698097}
\bibinfo{author}{\bibfnamefont{O.}~\bibnamefont{Scholten}},
  \emph{\bibinfo{title}{Mgmr3d}} (\bibinfo{year}{2023}),
  \urlprefix\url{https://doi.org/10.5281/zenodo.7698097}.

\end{thebibliography}
\end{document}